\documentclass[11pt,a4paper]{article}
\pdfoutput=1

\usepackage{jheppub}
\usepackage{latexsym}
\usepackage{multirow}
\usepackage{color}
\usepackage[usenames,dvipsnames,table]{xcolor}
\usepackage{graphicx}
\usepackage{epsfig}  
\usepackage{epsf}    
\usepackage{dcolumn}
\usepackage{bm}
\usepackage{dcolumn}
\usepackage{textcomp}
\usepackage{float}
\usepackage{subfig}
\usepackage{hypcap}
\usepackage[]{hyperref}
\usepackage{makecell}
\usepackage{color}
\usepackage{pifont}
\usepackage{appendix}

\hypersetup{
  bookmarks=true,         
  unicode=false,          
  pdftoolbar=true,        
 pdfmenubar=true,        
 pdffitwindow=true,     
 pdfstartview={FitH},    
 pdfsubject={Neutrino Oscillations Phenomenology},   
 pdfnewwindow=true,      
 pdfcreator={RevTeX},
 colorlinks=true,       
 linkcolor=red,          
 citecolor=blue,        
 filecolor=black,      
 urlcolor=blue,           
  }

\def\anu{{\bar\nu}}

\newcommand{\beq}{\begin{equation}}
\newcommand{\eeq}{\end{equation}}
\newcommand{\beqa}{\begin{eqnarray}}
\newcommand{\eeqa}{\end{eqnarray}}

\newcommand{\tx}{{\theta_{12}}}
\newcommand{\ty}{{\theta_{13}}}
\newcommand{\tz}{{\theta_{23}}}

\newcommand{\dcp}{\delta_{\mathrm{CP}}}

\newcommand{\pmue}{P(\nu_\mu \rightarrow \nu_e)}

\newcommand{\dxx}{\Delta\chi^2}
\newcommand{\dxxmin}{\Delta\chi^2_{\textrm{min}}}

\newcommand{\sig}{$\sigma~$}
\newcommand{\dmm}{\Delta m^2_{\mu\mu}}

\preprint{IP/BBSR/2014-9}

\title{Probing Neutrino Oscillation Parameters using High Power Superbeam from ESS}

\author[a]{Sanjib Kumar Agarwalla,}
\author[b,c]{~Sandhya Choubey,} 
{\author[b]{~Suprabh Prakash$\,$}

\affiliation[a]{Institute of Physics, Sachivalaya Marg, Sainik School Post, Bhubaneswar 751005, India}
\affiliation[b]{Harish-Chandra Research Institute, Chhatnag Road, Jhunsi, Allahabad 211019, India}
\affiliation[c]{Department of Theoretical Physics,
School of Engineering Sciences, KTH Royal Institute of Technology, AlbaNova University Center, 106 91 Stockholm, Sweden}

\emailAdd{sanjib@iopb.res.in}
\emailAdd{suprabhprakash@hri.res.in}
\emailAdd{sandhya@hri.res.in}

\abstract
{A high-power neutrino superbeam experiment at the ESS 
facility has been proposed such that the source-detector distance 
falls at the second oscillation maximum, giving very good sensitivity 
towards establishing CP violation. In this work, we 
explore the comparative physics reach of the experiment in terms of 
leptonic CP-violation, precision on atmospheric parameters,
non-maximal $\tz$, and its octant for a variety of choices for the baselines. 
We also vary the neutrino vs. the anti-neutrino running time for the 
beam, and study its impact on the physics goals of the experiment. 
We find that for the determination of CP violation, 540 km baseline 
with 7 years of $\nu$ and 3 years of $\bar\nu$ 
($7\nu+3\bar\nu$) run-plan performs the best and one expects  
a $5\sigma$ sensitivity to CP violation for 
48\% of true values of $\delta_{\rm CP}$. 
The projected reach for the 
200 km baseline with $7\nu+3\anu$ run-plan
is somewhat worse with $5\sigma$ sensitivity for 
34\% of true values of $\delta_{\rm CP}$. On the other hand, for 
the discovery of a non-maximal $\theta_{23}$ and its octant, 
the 200 km baseline option with 
$7\nu+3\bar\nu$ run-plan 
performs significantly better than the other baselines. 
A $5\sigma$ determination of a non-maximal $\tz$
can be made if the true value of $\sin^2\theta_{23} \lesssim 0.45$ or  
$\sin^2\theta_{23} \gtrsim 0.57$.
The octant of $\theta_{23}$ could be resolved at $5\sigma$ 
if the true value of $\sin^2\theta_{23} \lesssim 0.43$ or $\gtrsim 0.59$, 
irrespective of $\delta_{\rm CP}$.
}

\keywords{Neutrino Oscillation, ESS Facility, Long-baseline, CP Violation, Precision, Octant of $\theta_{23}$}
\arxivnumber{1406.2219}

\begin{document}
\maketitle
\flushbottom

\section{Introduction and Motivation}
\label{introduction}

The current explosion of activity in hunting for signals of physics beyond the Standard Model
of particle physics received tremendous boost with the widely confirmed claim that neutrinos 
have mass \cite{Agashe:2014kda}. The credit goes to the pioneering world-class experiments 
involving neutrinos from the Sun \cite{Cleveland:1998nv,Altmann:2005ix,Hosaka:2005um,Ahmad:2002jz,Aharmim:2008kc,Aharmim:2009gd,Arpesella:2008mt}, the Earth's atmosphere \cite{Fukuda:1998mi,Ashie:2005ik}, nuclear reactors \cite{Araki:2004mb,:2008ee,An:2012eh,An:2012bu,Ahn:2012nd,Abe:2011fz,Abe:2012tg}, and accelerators \cite{Ahn:2006zza,Adamson:2008zt,Adamson:2011qu,Adamson:2013ue,Abe:2011sj,Abe:2013xua}
which have established the phenomenon of neutrino flavor oscillations \cite{Pontecorvo:1967fh,Gribov:1968kq,Bilenky:2014eza} on a strong footing.
This immediately demands that neutrinos have mass and they mix with each other, providing an exclusive evidence for physics beyond the 
Standard Model.

With the recent discovery of the last unknown neutrino mixing angle $\theta_{13}$ 
\cite{An:2013zwz,An:2012bu,Abe:2012tg,Ahn:2012nd,An:2012eh,Abe:2011fz},
the focus has now shifted towards the determination of the remaining unknown parameters of the 
three generation neutrino flavor oscillation paradigm. These include the 
neutrino mass ordering, discovery of CP violation and measurement of the 
CP phase $\delta_{\rm CP}$ in the neutrino sector, and finally determination of the 
deviation of the mixing angle $\theta_{23}$ from maximal and its octant. 
Various experimental proposals have been put forth to nail these remaining 
parameters of the neutrino mass matrix. 
Measurement of non-zero $\theta_{13}$ has opened up the chances of 
determining the neutrino mass ordering, CP violation, as well as the octant of $\theta_{23}$.  
In particular, the relatively large value of $\theta_{13}$ has ensured that 
the neutrino mass ordering, {\it aka}, the neutrino mass hierarchy, could be determined  
to a rather high statistical significance in the next-generation proposed 
atmospheric \cite{Akhmedov:1998ui,Akhmedov:1998xq,Chizhov:1999az,Banuls:2001zn,Gandhi:2004md,Barger:2012fx}, 
long-baseline \cite{Pascoli:2013wca,Blennow:2013oma,::2013kaa,Agarwalla:2014fva}, 
and medium-baseline reactor experiments \cite{Li:2014qca,RENO-50}. 
The determination of the deviation of $\theta_{23}$ from its maximal value 
and its octant can also be studied in a variety of proposed long-baseline and 
atmospheric neutrino experiments \cite{Antusch:2004yx,Minakata:2004pg,GonzalezGarcia:2004cu,Choudhury:2004sv,
Choubey:2005zy,Indumathi:2006gr,Kajita:2006bt,Hagiwara:2006nn,Samanta:2010xm,Choubey:2013xqa,Chatterjee:2013qus,Agarwalla:2013ju}. 
The chances of exploring CP violation\footnote{For a detailed discussion on the CP violation discovery potential 
of T2K and NO$\nu$A, see for example \cite{Huber:2009cw,Agarwalla:2012bv,Machado:2013kya,Ghosh:2014dba}.} 
in a given experiment depend on how well one can probe the CP asymmetry $A_{CP}$ which is defined as 
$(P-\bar{P})/(P+\bar{P})$ where $P(\bar{P})$ are the neutrino (anti-neutrino) probability \cite{Dick:1999ed, Donini:1999jc,Minakata:2012ue}. 
New experiments with more powerful beams and bigger detectors have been proposed to enhance the CP discovery potential. 

There has been a proposal to extend the European Spallation Source (ESS) program 
to include production of a high intensity neutrino beam, which is being called the 
European Spallation Source Neutrino Super Beam (ESS$\nu$SB) \cite{Baussan:2012cw,Baussan:2013zcy}. 
Since the neutrino beam is expected to have energies in the few 100s of MeV regime, 
the proposed detector is a 500 kt MEMPHYS \cite{Agostino:2012fd,Campagne:2006yx} 
type water Cherenkov detector. The collaboration aims to gain from the R\&D already performed for the 
SPL beam proposed at CERN and the MEMPHYS detector proposed at 
Frejus. The optimization of the peak beam energy and baseline of the experiment have  
been studied in \cite{Baussan:2013zcy} in terms of the  CP violation discovery reach of this set-up. 
The choice of peak beam energy of 0.22 GeV and baseline 500 km for this experimental proposal  
returns a $3\sigma$ CP violation discovery potential for almost 70\% of $\dcp$(true) 
values \cite{Baussan:2013zcy}. In this paper, we focus on the octant of $\theta_{23}$ and its 
deviation from maximal mixing for a superbeam experiment using a 
ESS$\nu$SB type beam and MEMPHYS type detector. We will use the 
ESS$\nu$SB corresponding to 2 GeV protons and consider 500 kt of detector 
mass for the water Cherenkov far detector and the optimize the experimental set-up  
taking various possibilities for the baseline of the experiment as well for a different 
run-time fractions of the beam in the neutrino and anti-neutrino modes. 

There remains some tension between the best-fit 
$\theta_{23}$ obtained from the analysis of the MINOS data \cite{Adamson:2014vgd}
with the best-fit $\theta_{23}$ coming from the 
analysis of the Super-Kamiokande (SK)
atmospheric neutrino data \cite{Himmel:2013jva},
as well as the latest data from the T2K experiment \cite{Abe:2014ugx}.
While the MINOS combined long baseline and atmospheric 
neutrino data yield the best-fit $\sin^2\theta_{23}=0.41(0.61)$
for the lower(higher) octant with a slight preference for the lower octant, 
SK atmospheric data gives the best-fit at $\sin^2\theta_{23}=0.6$
for both normal hierarchy (NH) and inverted hierarchy (IH),
and T2K gives the best-fit at $\sin^2\theta_{23}=0.514(0.511)$
for NH(IH). 
The current global fits of the existing world neutrino data 
by different groups too 
give conflicting values for 
the best-fit $\sin^2\theta_{23}$. While the analysis in 
\cite{Capozzi:2013csa} gives the 
best-fit $\sin^2\theta_{23}=0.437(0.455)$ for NH(IH),  
the analysis in \cite{Forero:2014bxa} gives the best-fit 
$\sin^2\theta_{23}=0.57$ for both NH and IH. 
In particular, different data sets and different analyses 
give conflicting answers to the question on whether 
$\theta_{23}$ is maximal. While the preliminary results
from T2K indicates near maximal mixing, SK and MINOS data 
disfavor maximal mixing at slightly over $1\sigma$.
On the other hand, the global fits are all inconsistent with 
maximal $\theta_{23}$ at less than $1\sigma$ (if we do not assume 
any knowledge on the mass hierarchy) and have conflicting 
trends on its octant (irrespective of the hierarchy). 
Though the tension on the value of $\theta_{23}$ and its octant 
between the different data sets and 
analyses are not statistically significant, nonetheless they are there, 
and need to be resolved at the  
on-going and next-generation neutrino facilities. In 
addition to determining the value of $\sin^22\theta_{23}$, we 
would also like to determine the $\theta_{23}$ octant, in case $\theta_{23}$ is 
found to be indeed non-maximal.  
The prospects of determining the octant of $\theta_{23}$ has been 
studied before in \cite{Antusch:2004yx,Minakata:2004pg,GonzalezGarcia:2004cu,Choudhury:2004sv,
Choubey:2005zy,Indumathi:2006gr,Kajita:2006bt,Hagiwara:2006nn,Samanta:2010xm,Choubey:2013xqa,Chatterjee:2013qus} 
using atmospheric neutrinos and accelerator-based neutrinos beams, 
and in \cite{Minakata:2002jv,Hiraide:2006vh} using reactor neutrinos.
We checked that the combined data from present generation long-baseline
experiments, T2K and NO$\nu$A can establish a non-maximal $\tz$ only 
if $\sin^2\theta_{23}$(true)$ \lesssim 0.45$ and $\gtrsim 0.57$ at $3\sigma$.
The same data can settle the octant of $\theta_{23}$ at $2\sigma$
provided $\sin^2\theta_{23}$(true) $ \lesssim 0.43$ and $\gtrsim 0.58$ 
irrespective of the value of $\delta_{\rm CP}$ \cite{Agarwalla:2013ju}.
Therefore, it is pertinent to ask whether the next-generation
long-baseline experiments can improve these bounds further.
Prospects of determining the octant of $\theta_{23}$ has been studied 
in \cite{Agarwalla:2013hma,Barger:2013rha,Barger:2014dfa} for the 
Long Baseline Neutrino Experiment (LBNE) proposal in the US, and 
for the Long Baseline Neutrino Oscillation (LBNO) experimental proposal 
in Europe in \cite{Agarwalla:2013hma,Ghosh:2013pfa}.
With the help of T2K and NO$\nu$A data, LBNE10 can determine
the octant of $\tz$ at 3$\sigma$ if $\sin^2\theta_{23}$(true) $\lesssim$ 0.44 and 
$\gtrsim$ 0.59 for any $\delta_{\rm CP}$ \cite{Agarwalla:2013hma}. 
The LBNO proposal with a 10 kt LArTPC can do this job if 
$\sin^2\theta_{23}$(true) $\lesssim$ 0.45 and $\gtrsim$ 0.58 \cite{Agarwalla:2013hma}.

In this work, we study in detail the achievable precision on
the atmospheric parameters and the prospects of determining the deviation of 
$\theta_{23}$ from maximal and its correct octant with the ESS$\nu$SB 
experiment. We consider various baseline and run-plan possibilities 
for this set-up and optimize them for best reach for $\theta_{23}$ octant 
such that the CP violation discovery reach of the experiment is not 
significantly compromised.
The paper is organized as follows. In section \ref{exp_spec}, we briefly
describe the ESS$\nu$SB proposal from the phenomenological viewpoint.
In section \ref{simulations}, we give the details of the simulation procedure.
In section \ref{results}, we describe the results we obtain regarding the 
sensitivities of the ESS$\nu$SB  set-up. Finally, in section \ref{sec:conclusions}, we 
give our conclusions.

\section{Experimental Specifications}
\label{exp_spec}

In this section we briefly describe the super beam set-up that we have considered in this 
study. The ESS project is envisaged as a major European facility providing 
slow neutrons for research as well as the industry. It is projected to start 
operation by 2019. 
The ESS$\nu$SB proposal is an extension of the original ESS facility to 
generate an intense neutrino beam for neutrino oscillation studies. The proposal is to 
use the 5 MW ESS proton driver with 
2 GeV protons, to produce a high intensity neutrino superbeam simultaneously 
along with the 
spallation neutrons, without compromising on the number of spallation neutrons. 
This dual purpose machine would result in considerable reduction of costs 
in contrast to the building of two separate proton drivers, one for neutrons 
and another for neutrinos. The proton driver could later be used as a part of the 
neutrino factory, if and when one is built. Detailed feasibility studies for this 
dual purpose machine is underway. We refer the readers to \cite{Baussan:2013zcy} for a detailed 
discussion on the accelerator, target station and the beam line being 
discussed for this proposal. 
While the proposed proton energy for the ESS facility is 2 GeV, the energy 
of the protons could be increased up to 3 GeV. 
The expected neutrino flux for this facility has been calculated for proton 
energy of 2 GeV and $2.7\times 10^{23}$ protons on target per year, 
corresponding to 5 MW power for the beam. For the other proton energies 
of 2.5 GeV and 3 GeV, the neutrino flux is calculated by 
keeping the power of the beam fixed at 5 MW. 
In this paper, we use the neutrino fluxes corresponding to the 2 GeV proton beam 
and $2.7\times 10^{23}$ protons on target per year \cite{Baussan:2013zcy}.

The on-axis neutrino flux for the 2 GeV protons on target peaks at 0.22 GeV.
Hence, megaton class water Cherenkov detector has been proposed as the 
default detector option for this set-up. At these energies, the detection 
cross-section is dominated by quasi-elastic scattering. 
We have used the GLoBES software \cite{Huber:2004ka,Huber:2007ji}
to simulate the ESS$\nu$SB set-up. We obtain the fluxes from \cite{enrique}
and consider the properties of the MEMPHYS 
detector \cite{Agostino:2012fd,luca} to simulate the events.
We take the fiducial mass of the detector to be 500 kt and 
a total run-time of 10 years.

For the peak neutrino energy of 0.22 GeV obtained for the 2 GeV protons on target, the 
first oscillation maximum corresponds to 180 km while the second oscillation maximum 
comes at 540 km. The possible detector locations are discussed in \cite{Baussan:2013zcy}.
Existing mines in Sweden where the detector can be housed are at distances of about 
260 km (Oskarshamn), 360 km (Zinkgruvan), 540 km (Garpenberg) and 1090 km (Kristineberg) 
from the ESS site, which is in Lund. The study in \cite{Baussan:2013zcy} uses the mine 
location at Garpenberg to place the detector, giving a baseline of 540 km which corresponds 
to the second oscillation maximum, well suited for CP violation discovery \cite{Coloma:2011pg}. 
The study shows that the CP violation discovery can be achieved for up to 50\% values 
of $\dcp$(true) at more than $5\sigma$. In what follows, we optimize the baseline 
for the deviation of $\theta_{23}$ from maximal and its octant, without severely
compromising the sensitivity to CP violation.
The number of events that we get for the set-up described above is shown in 
Table \ref{tab:events}. It can be seen that event numbers in 
Table \ref{tab:events} have a good match with Table 3 of \cite{Baussan:2013zcy}.
 
\begin{table}[h]
\centering
{\footnotesize
\begin{tabular}{|r|c|c|c|c|c|c|}
 \hline
Set-up
&$\nu_{e} (\bar{\nu}_{e})$
&$\nu_{\mu} (\bar{\nu}_{\mu})$
&$\nu_{e}$
&$\bar{\nu}_{e}$
&NC
&$\bar{\nu}_{\mu}(\nu_{\mu})\rightarrow\bar{\nu}_{e}(\nu_{e})$
\\
&signal
&miss-ID
&intrinsic
&intrinsic
&
&wrong-sign contamination
\\
\hline
360 km ($\nu$ run)
&304
&10
&75
&0.08
&25
&1.0
\\
($\anu$ run)
&244
&6
&3
&53
&15
&11
\\
\hline
540 km ($\nu$ run)
&197
&5
&34
&0.04
&11
&0.7
\\
($\anu$ run)
&164
&3
&1
&24
&7
&7
\\
\hline
\end{tabular}
}
\caption{\label{tab:events} \footnotesize{Signal and background events for 
the ESS$\nu$SB set-up with a 360 km baseline and a 540 km baseline. Both
$\nu$ and $\anu$ events are shown.
To generate these numbers, we used the following values
of the neutrino oscillation parameters: $\Delta m^2_{21} = 7.5\times10^{-5}\rm{eV^2}$,
$\Delta m^2_{31} = 2.47\times10^{-5}\rm{eV^2}$, $\sin^2\tx=0.3$, $\sin^22\ty=0.087$,
$\sin^2\tz=0.415$ and $\dcp=0$. These values are the same as that used to generate
Table 3 of \cite{Baussan:2013zcy}.}}
\end{table}

\section{Oscillation Probability and Simulation Details}
\label{simulations}

Here, we focus on the relevant oscillation channels and simulation
methods which go in estimating the final results.
\subsection{$\tz$-dependence in the disappearance and appearance channels}
The precision measurement of the mixing angle $\theta_{23}$ in 
long-baseline experiments comes from the disappearance 
channel. This channel depends on 
the survival probability for muon neutrinos, which 
in the approximation that $\Delta m^2_{21}=0$ is given as \cite{Choubey:2005zy}
\beqa
P(\nu_\mu\to \nu_\mu) \approx 1 &-& \sin^2\theta_{13}^M\sin^22\theta_{23}\sin^2
\frac{[(\Delta m_{31}^2 + A)-(\Delta m_{31}^2)^M]L}{8E} \nonumber \\
&-&\cos^2\theta_{13}^M\sin^22\theta_{23}\sin^2
\frac{[(\Delta m_{31}^2 + A)+(\Delta m_{31}^2)^M]L}{8E} \nonumber \\
&-&\sin^22\theta_{13}^M\sin^4\theta_{23}\sin^2
\frac{(\Delta m_{31}^2)^M L}{4E}\,,
\label{eq:pmm}
\eeqa 
where $\theta_{13}^M$ and $(\Delta m_{31}^2)^M$ are the 
mixing angle $\theta_{13}$ and $\Delta m_{31}^2$ in matter 
and
$A$ is the Wolfenstein matter term \cite{msw1} and is given by
$A ({\rm eV}^2) = 0.76 \times 10^{-4} \rho \ ({\rm g/cm^3}) E ({\rm GeV})$.  
The disappearance data through its sensitivity to $\sin^22\tz$
as seen in the leading first term in Eq. (\ref{eq:pmm})
provides stringent constraint. This provides a 
powerful tool for testing a maximal $\theta_{23}$ against a non-maximal 
one. However, 
the leading first term does not depend on the octant of $\theta_{23}$.
This dependence comes only at the sub-leading level from the 
third term in Eq. (\ref{eq:pmm}), which becomes relevant only 
when matter effects are very large to push $\sin^2\theta_{13}^M$ 
close to resonance. Since the ESS$\nu$SB set-up involves very 
low neutrino energies and short baselines, the disappearance 
channel would provide almost no octant sensitivity and if $\theta_{23}$ 
was indeed non-maximal, it would give 
narrow allowed-regions in both the lower and the higher octant of $\tz$.

The octant sensitivity of long baseline experiments come predominantly  
from the electron appearance channel which depends on the $\pmue$
transition probability. Since this channel also gives sensitivity to 
CP violation for non-zero $\Delta m_{21}^2$, we give here 
the $\nu_{\mu} \to \nu_{e}$ oscillation probability in matter, 
expanded perturbatively in 
$\alpha(= \Delta m_{21}^2/\Delta m_{31}^2)$ and 
$\sin\theta_{13}$, keeping up to the second order terms in 
these small parameters 
\cite{Freund:2001pn,Akhmedov:2004ny,Cervera:2000kp} 
\beqa
\nonumber P\left(\nu_{\mu}\to \nu_{e}\right) &\sim& P_{\mu e} = 
\sin^2 2\ty\sin^2\tz
\frac{\sin^2\hat{\Delta}(1 - \hat{A})}{(1 - \hat{A})^2}  \nonumber \\ 
&& 
+\alpha\cos\ty\sin2\tx\sin2\ty\sin2\tz\cos(\hat{\Delta}+\dcp)\frac{\sin\hat{\Delta}\hat{A}}{\hat{A}}
\frac{\sin\hat{\Delta}(1-\hat{A})}{1-\hat{A}} \nonumber \\
&&
 + \alpha^2\sin^22\tx\cos^2\ty\cos^2\tz\frac{\sin^2\hat{\Delta}\hat{A}}{\hat{A}^2}  
\label{pmue}
\eeqa
where $\hat{\Delta}= \Delta m_{31}^2 L/4E$ and $\hat{A}= A/\Delta m_{31}^2$ 
are dimensionless parameters. The leading first term in Eq. (\ref{pmue}) 
depends on the octant of $\theta_{23}$. Octant dependence comes 
also from the third term, however this term is suppressed at second 
order in $\alpha$. 
The $\dcp$ 
dependence comes only in the second term which goes as 
$\sin2\tz$. 
However, it was shown in \cite{Agarwalla:2013ju} that the 
presence of the $\dcp$ term in the probability brings in 
a $\dcp-\theta_{23}$ degeneracy which can be alleviated 
only through a balanced run of the experiment between the 
neutrino and anti-neutrino channels. 

The approximate expressions in this section is given only for illustration. 
Our numerical analysis is done using the full three-generation oscillation 
probabilities. For the analysis performed in this paper, we simulate
predicted events at the following true values of the oscillation parameters: 
$\sin^22\theta_{13}$ = 0.089, 
$\Delta m_{21}^2$ = 7.5 $\times 10^{-5}$ eV$^2$, 
$\sin^2\theta_{12}$ = 0.3, while the values for $\theta_{23}$ 
and $\dcp$ are varied within their allowed ranges. 
We take the true value of atmospheric splitting to be
$\Delta m_{\mu\mu}^2$ = $\pm$ $2.4 \times 10^{-3}$ eV$^2$
where +ve (-ve) sign is for NH (IH).
The relation between $\Delta m_{\mu\mu}^2$ and 
$\Delta m_{31}^2$ has been taken 
from \cite{Nunokawa:2005nx,deGouvea:2005hk}.
Our assumptions for the systematic uncertainties considered 
are as follows. For the appearance channel, we take 10\% signal
normalization error and 25\% background normalization error.
For the disappearance events, we take 5\% signal
normalization error and 10\% background normalization error.
For both types of events, a 0.01\% energy calibration error has
been assumed. These `simulated events' is then fitted by means of a $\chi^2$ 
to determine the sensitivity of the experiment to the different 
performance indicators. We use the following definition of
$\chi^2$:
\beq
\chi^2= min_{\xi_s, \xi_b}\left[2\sum^{n}_{i=1}
(\tilde{y}_{i}-x_{i} - x_{i} \ln \frac{\tilde{y}_{i}}{x_{i}}) +
\xi_s^2 + \xi_b^2\right ]~,
\label{eq:chipull}
\eeq
where $n$ is the total number of bins and
\beq
\tilde{y}_{i}(\{\omega\},\{\xi_s, \xi_b\}) = N^{th}_i(\{\omega\}) \left[
1+ \pi^s \xi_s \right] +
N^{b}_i \left[1+ \pi^b \xi_b \right]~.
\label{eq:rth}
\eeq

Above, $N^{th}_i(\{\omega\})$ is the predicted number of events
in the $i$-th energy bin for a set of oscillation parameters $\omega$ 
and $N_i^b$ are the number of background events in bin $i$. 
The quantities $\pi^s$ and $\pi^b$ in Eq.~\ref{eq:rth} are the systematical errors on
signals and backgrounds respectively. The quantities $\xi_s$
and $\xi_b$ are the pulls due to the systematical error on signal and
background respectively. $x_{i}$ is the predicted event rates corresponding to
the $i$-th energy bin, consisting of signal and backgrounds. $\chi^2$ corresponding
to all the channels defined in the experiment are calculated and summed over.
Measurements of oscillation parameters available from other experiments 
are incorporated through Gaussian priors.
\beq
\chi^2_{\textrm{total}} = \sum_{j=1}^{c} \chi^2_j + \chi^2_{\textrm{prior}}
\eeq where c is the total number of channels. Finally,
$\chi^2_{\textrm{total}}$ is marginalized in the fit over the allowed
ranges in the oscillation parameters to find $\dxxmin$. More details
of $\chi^2$ definition, as given in Eqs. \ref{eq:chipull} and \ref{eq:rth},
can be found in \cite{Fogli:2002au, Huber:2002mx}.

\subsection{Numerical procedure}

\underline{Leptonic CP-violation}:
To evaluate the sensitivity to leptonic CP-violation, we follow the 
following approach. We first assume a true value of $\dcp$ lying in the
allowed range of $[-180^\circ,180^\circ]$. The event spectrum assuming this true
$\dcp$ is calculated and is labeled as {\it predicted} event spectra. 
We then calculate the various {\it theoretical}
event spectra assuming the test $\dcp$ to be the CP-conserving values
0 or $\pi$ and by varying the other oscillation parameters in their 
$\pm2\sigma$ range (the solar parameters are not varied) except $\tz$ 
which is varied in the $\pm3\sigma$ range.
We add prior on $\sin^22\ty$ ($\sigma=5\%$) as expected after the full run of 
Daya Bay \cite{dayabay_NF12}. We use the
software GLoBES to calculate the $\dxx$ between each set of predicted and
theoretical events. The smallest of all such $\dxx$: $\dxxmin$ is considered.
The results are shown by plotting $\dxxmin$ as a function of assumed 
true value in the range $[-180^\circ,180^\circ]$.
\\
\underline{Precision on $\dmm$ and $\sin^2\tz$}: We simulate the {\it predicted}
events due to a true value of $\dmm$. For generating 
the {\it theoretical} spectrum, values of $\dmm$ in the $\pm2\sigma$
range around the central true value are chosen. We marginalize over rest
of the oscillation parameters including hierarchy in order to calculate
the $\dxxmin$. Similar procedure is followed in the case of $\sin^2\tz$,
with the exception that for non-maximal true values of $\tz$, we confine
the test range to be in the true octant only.
\\
\underline{Sensitivity to maximal vs. non-maximal $\theta_{23}$}:
We consider true $\sin^2\tz$ values in the allowed 3\sig range
and calculate events, thus simulating the true events. This is then contrasted 
with {\it theoretical} event spectra assuming the test $\sin^2\tz$
to be 0.5. Rest of the oscillation parameters, including hierarchy, 
are marginalized to obtain the least $\dxx$. This procedure is done 
for a fixed true $\dcp$ value of 0 and normal mass hierarchy. 
\\
\underline{Sensitivity to Octant of $\theta_{23}$}:
To calculate the sensitivity to the octant of $\tz$, the following 
approach is taken. We take a true value of $\sin^2\tz$ lying in the 
lower octant. The other known oscillation parameters are kept
at their best-fit values. Various test $\sin^2\tz$ values are
taken in the higher octant. Test values for other oscillation
parameters are varied in the $\pm2\sigma$ range. We marginalize
over the hierarchies. $\dxx$ values 
between the true and test cases are calculated and the least of 
all such values: $\dxxmin$ is considered. This is repeated for 
a true $\sin^2\tz$ lying in the higher octant, but this time
the test values of $\sin^2\tz$ are considered from the lower octant
only. This is done for both NH and IH as true choice and various values 
of $\dcp$(true) in $[-180^\circ,180^\circ]$.

\section{Results}
\label{results}

In this section, we report our findings regarding the leptonic CP-violation,
achievable precision on atmospheric parameters, 
non-maximality of $\tz$ and its octant for the proposed ESS$\nu$SB set-up.

\subsection{Discovery of leptonic CP-violation}

\begin{figure}[h]
\centering
\includegraphics[width=0.49\textwidth]
{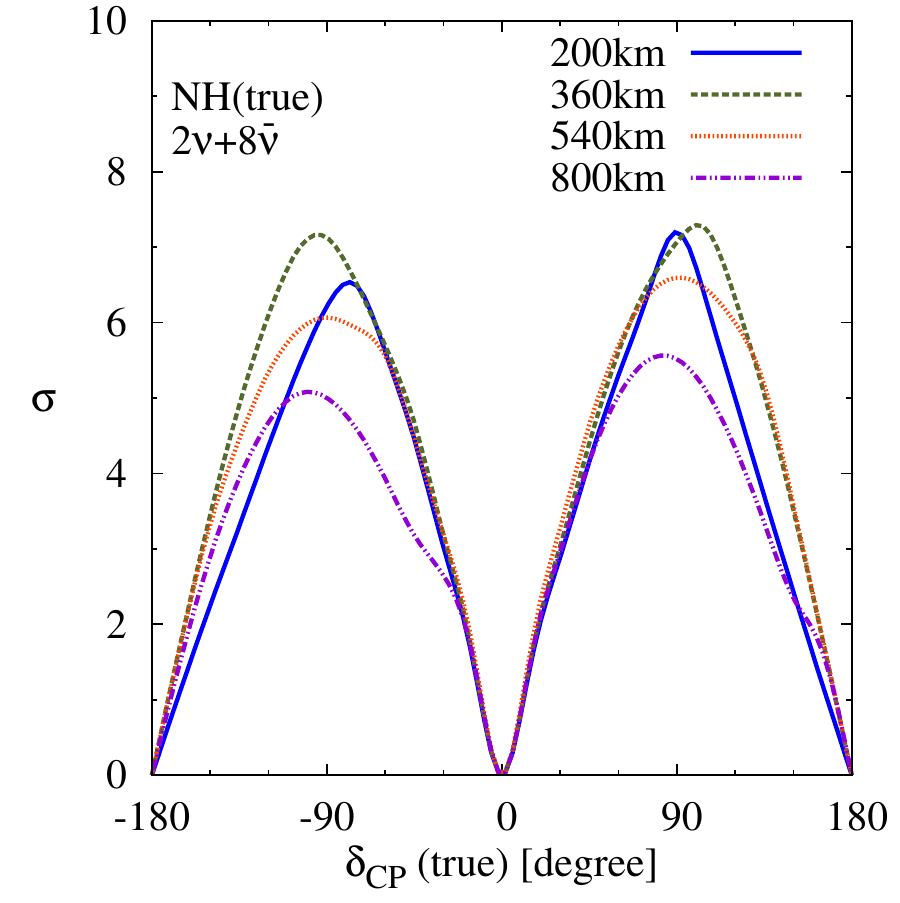}
\includegraphics[width=0.49\textwidth]
{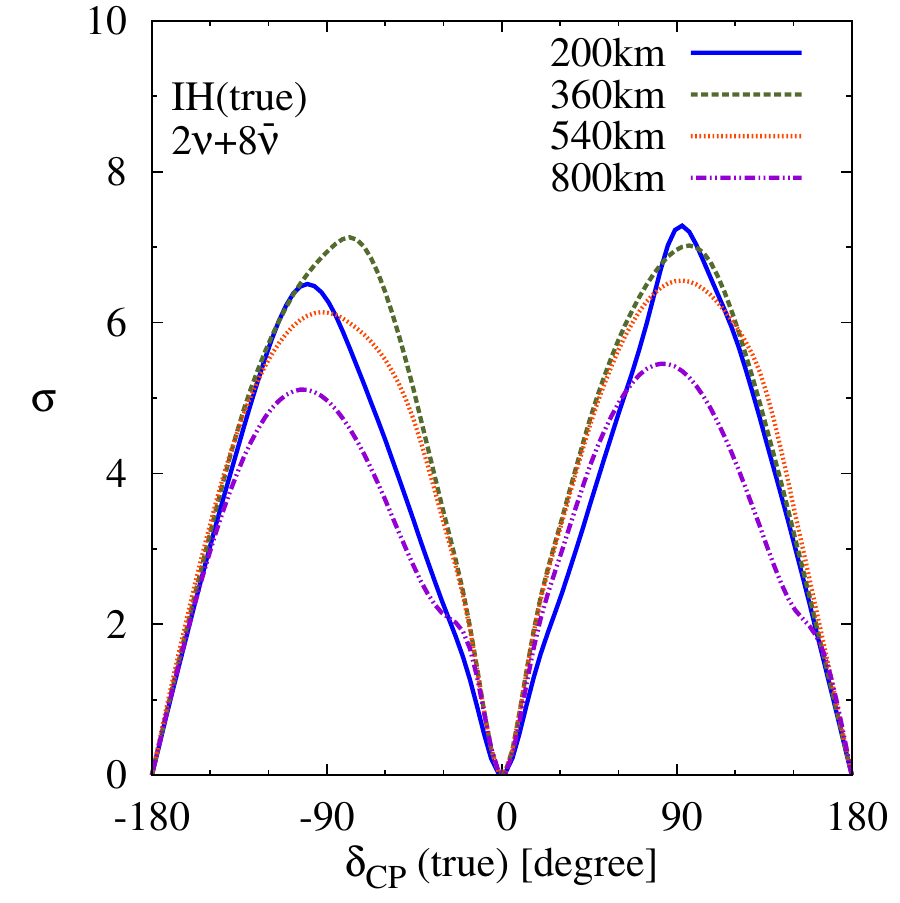}
\caption{\footnotesize{CP violation discovery potential 
(in $\sigma$) as a function of $\dcp$(true).
The left(right) panel assumes NH(IH) to be the true hierarchy. Baselines 
corresponding to 200 km, 360 km, 540 km and 800 km have been considered. 
The choice of run-plan is $2\nu+8\anu$ years of running. }}
\label{cpviolation_diffbaselines}
\end{figure}

We first show the results for the sensitivity of the ESS$\nu$SB set-up to CP violation.
We compare the sensitivity of the set-up for different possible baseline options. 
We have chosen the representative values of 200 km, 360 km, 540 km and 800 km which are 
the same as what has been considered in \cite{Baussan:2013zcy}. 
In Fig. \ref{cpviolation_diffbaselines}, we show the discovery reach towards CP violation 
for these prospective baselines.\footnote{It should be noted that for producing the results for CP violation,
the values of true oscillation parameters considered are the same as those
in Table \ref{tab:events}. While, these values are the same as those considered 
in \cite{Baussan:2013zcy}, they are different from what we have taken for producing 
other results in this paper.}
In the y-axis, we have plotted the 
confidence level (C.L.), (defined as $\sqrt{\dxxmin}$)
and in the x-axis we have plotted the true $\dcp$ values lying in the 
range $[-180^\circ,180^\circ]$. The left panel is assuming the NH to be the true
hierarchy while, in the right panel we have assumed IH to be the true
hierarchy. 
The run plan
considered here is two years of neutrino running followed by eight
years of anti-neutrino running ($2\nu+8\anu$), to match with the run plan 
assumed in the ESS$\nu$SB proposal \cite{Baussan:2013zcy}. 
In producing these plots, we have considered the test
hierarchy to be the same as the true one which implies that we have not
marginalized over hierarchies while calculating the $\dxx$. Note that 
the CP discovery reach results shown in \cite{Baussan:2013zcy} are obtained after 
marginalizing over the neutrino mass hierarchy. We have performed our 
analysis for the CP discovery reach both with and without marginalizing over 
the mass hierarchy and have presented the results for the fixed test hierarchy case.
The underlying 
justification for doing this is the fact that by the time this experiment comes
up, we may have a better understanding of the neutrino mass hierarchy. In addition, 
from the observation of atmospheric neutrino events in the 
500 kt water Cherenkov detector deployed for the ESS$\nu$SB set-up, 
$3\sigma$ to $6\sigma$ sensitivity to the mass hierarchy 
is expected, depending on the true value of 
$\sin^2\theta_{23}$. Here one assumes that 
the ESS$\nu$SB far detector will have similar features to the Hyper-Kamiokande 
proposal in Japan \cite{Abe:2011ts}. 
The impact of marginalization over the hierarchy is mainly in reducing somewhat the 
CP coverage for the $L=200$ km baseline option. For the other baselines, the impact 
of marginalizing over the test hierarchy is lower mainly because for these longer baselines 
the hierarchy degeneracy gets resolved via the ESS$\nu$SB set-up alone. 

Fig. \ref{cpviolation_diffbaselines} shows that our results for CP violation are in 
agreement with those in \cite{Baussan:2013zcy}. 
From the left panel of Fig. \ref{cpviolation_diffbaselines}, it can 
be seen that for the 200 km baseline, which is the smallest amongst the
four choices considered, a 3\sig C.L. 
evidence of CP violation is possible for 
60\% of $\dcp$(true), while a 32\% coverage is possible at
5\sig C.L. For the 540 km baseline, which shows the best sensitivities
among the four choices considered, discovery of CP violation 
at the 3\sig C.L. is expected to be possible for 
70\% of $\dcp$(true),
while a 5\sig significance is expected for 
45\% of $\dcp$(true).
Thus, we are led
to the conclusion that the 540 km choice is better-suited for the
discovery of CP-violation with this set-up than any other choice of
baseline. However, the CP violation discovery reach of the 360 km and 200 km 
baselines are only marginally lower. In particular, we note that if we have to 
change from the 540 km baseline to 200 km baseline, the CP coverage 
for CP violation discovery goes down only by $\sim$13\% (10\%) at the 5\sig (3\sig) C.L.

\begin{figure}[h]
\centering
\includegraphics[width=0.49\textwidth]
{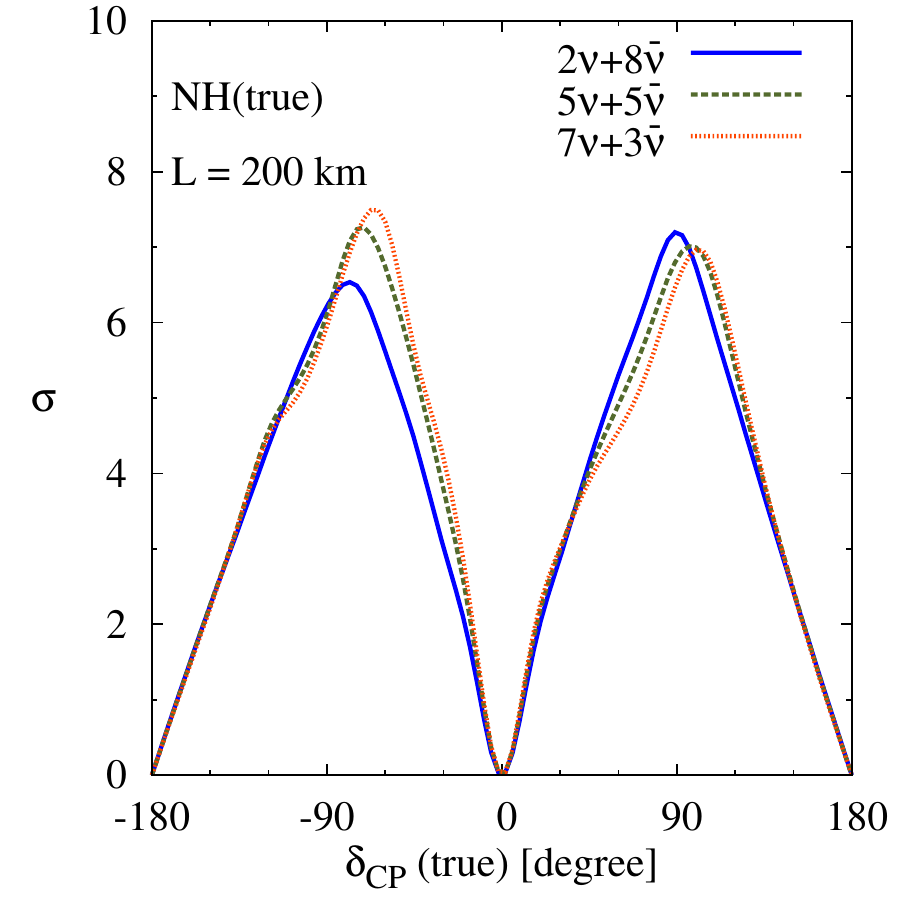}
\includegraphics[width=0.49\textwidth]
{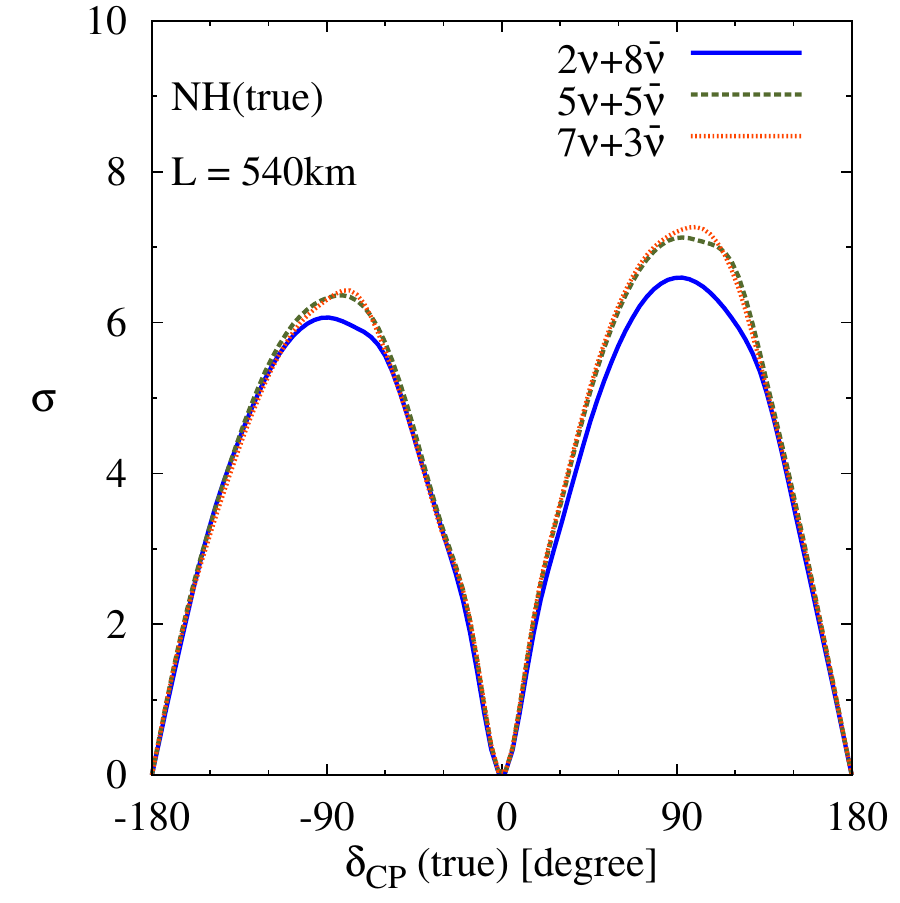}
\caption{\footnotesize{Statistical significance ($\sigma$) for CP violation
discovery potential as a function of $\dcp$(true).
NH has been assumed to be the true hierarchy. The left(right)
panel corresponds to the choice of 200 km (540 km) as the baseline. Results
for different run-plans corresponding to $2\nu+8\anu$, $5\nu+5\anu$
and $7\nu+3\anu$ years of running have been shown.}}
\label{cpviolation_diffruns}
\end{figure}

In \cite{Baussan:2013zcy}, the nominal choice for the 
neutrino vs. anti-neutrino run-plan for the ESS$\nu$SB was taken 
as $2\nu+8\anu$. The motivation behind this choice was to have 
similar number of events for both $\nu$ and $\anu$ running.
However, in order to explore this further, we calculate the sensitivity to
CP violation for different run-plans. We have taken three
cases: $2\nu+8\anu$, $5\nu+5\anu$, and $7\nu+3\anu$. 
The left (right) panel in Fig. \ref{cpviolation_diffruns} shows the projected 
CP discovery potential for the 200 km (540 km) baseline option, 
for different run-plans. From these plots, it can be seen that
at lower C.L., all the three run-plans have similar sensitivity.
However, at 5\sig C.L., the larger coverage in $\dcp$ comes with
$7\nu+3\anu$ running. While this holds true for both 200 km and 540 km, the 
effect is marginally more pronounced for the 200 km baseline option. 

\subsection{Precision on atmospheric parameters}

We now focus on the achievable precision on atmospheric parameters
with the proposed set-up. The precision\footnote{We define the relative 
1\sig error as 1/6th of the $\pm3\sigma$ variations around the
true choice.} is mainly governed by the 
$P(\nu_{\mu}\rightarrow\nu_{\mu})$ channel (see Eq. (\ref{eq:pmm})). 
Because of huge statistics
in this channel, we expect this set-up to pin down the atmospheric 
parameters to ultra-high precision. Indeed, this is the case as can
be seen from Table \ref{tab:precision}. Table \ref{tab:precision}
shows the relative 1\sig precision on $\dmm$ and $\sin^2\tz$ 
considering three different values of true $\sin^2\tz$. Here, we
have taken the baseline to be 200 km and the run-plan to be 
$7\nu+3\anu$.

\begin{table}[h]
\centering
{\footnotesize
\begin{tabular}{|r|c|c|c|}
 \hline
$\sin^2\tz$(true)
&0.4
&0.5
&0.6
\\
\hline
$\delta(\dmm)$
&0.24\%
&0.2\%
&0.22\%
\\
\hline
$\delta(\sin^2\tz)$
&1.12\%
&3.0\%
&0.8\%
\\
\hline
\end{tabular}
}
\caption{\label{tab:precision} \footnotesize{Relative 1\sig precision (1 dof) on
$\dmm$ and $\sin^2\tz$ considering three different values of 
true $\sin^2\tz$. Here, for all the cases, we consider the true value
of $\dmm$ to be $2.4\times10^{-3}\mathrm{eV^2}$. We consider NH as
the true hierarchy. We have considered the 200 km as the baseline 
and $7\nu+3\anu$ as the run-plan for generating these numbers.}}
\end{table}

It can be seen that around 0.2\% precision on the atmospheric mass splitting is achievable
which is a factor of $\sim5$ better than what can be achieved with combined data from 
T2K and NO$\nu$A \cite{Agarwalla:2013qfa}. While the precision on $\dmm$ is 
weakly-dependent on the true value of $\sin^2\tz$, the precision in $\sin^2\tz$ shows
a large dependence on its central value. We see that for $\sin^2\tz=0.5$, the precision 
is 3.0\%, while for $\sin^2\tz=0.6$, its 0.8\%. The precision in $\sin^2\tz$
is worst for the maximal mixing due to the fact that a large Jacobian 
is associated with transformation of the variable from $\sin^22\tz$
to $\sin^2\tz$ around the maximal mixing \cite{Minakata:2004pg}.

\subsection{Deviation from maximality}

\begin{figure}[h]
\centering
\includegraphics[width=0.49\textwidth]
{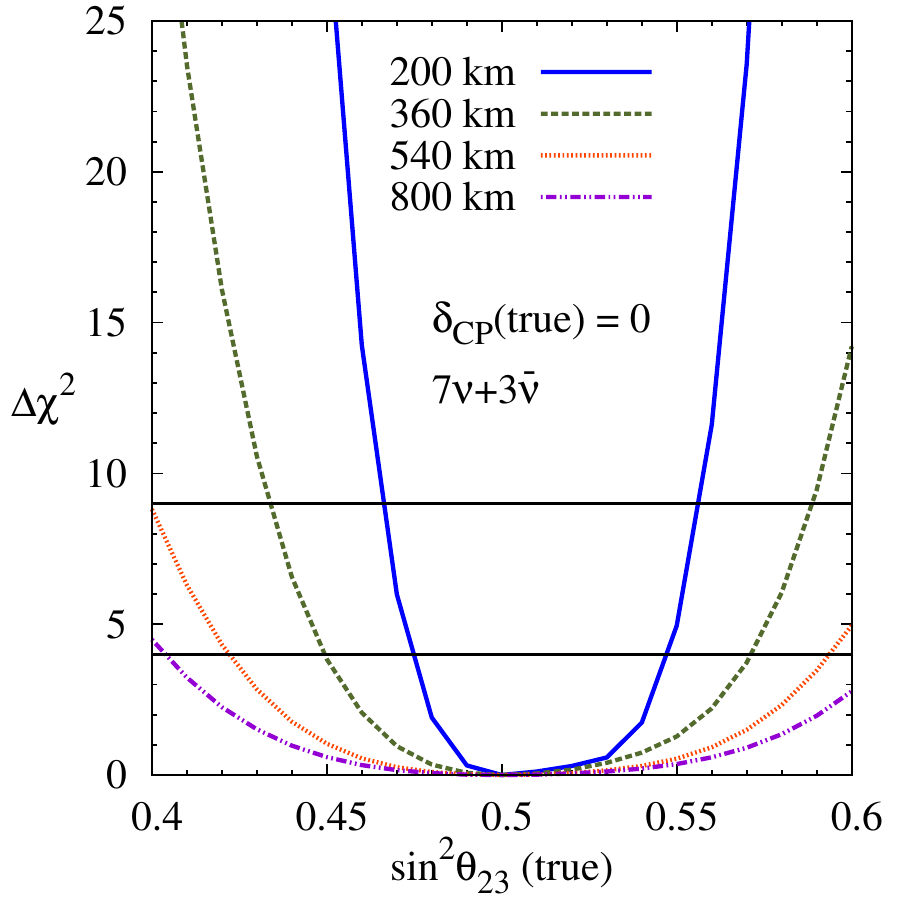}
\includegraphics[width=0.49\textwidth]
{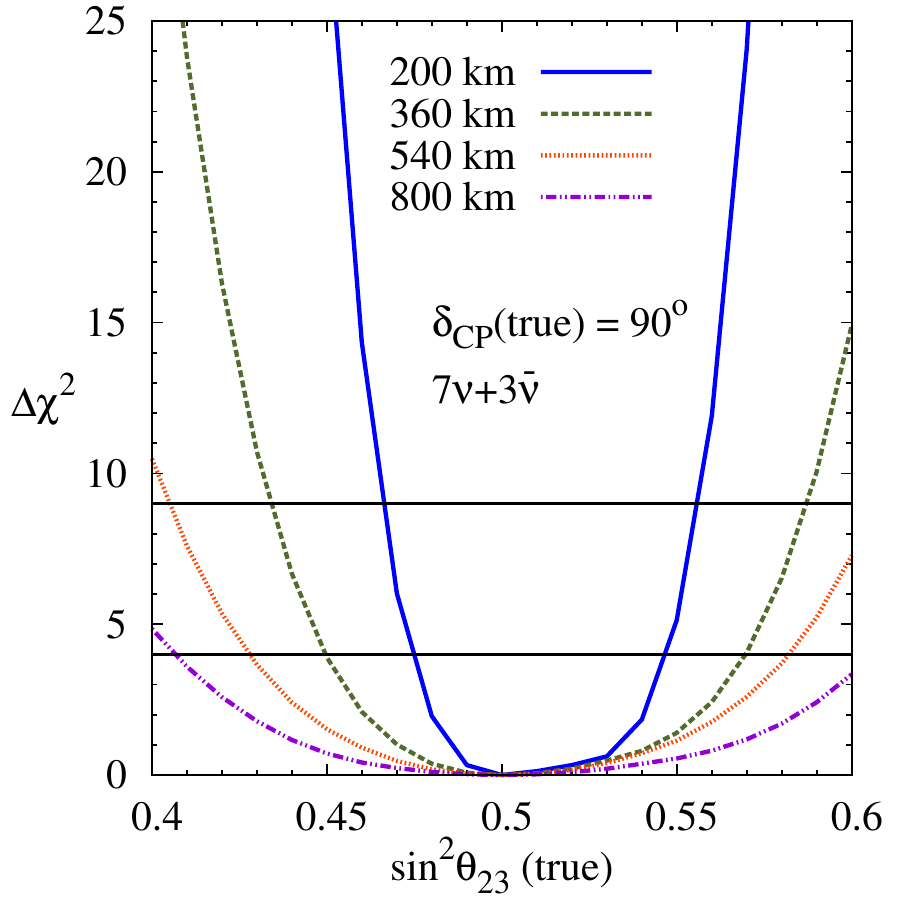}
\includegraphics[width=0.49\textwidth]
{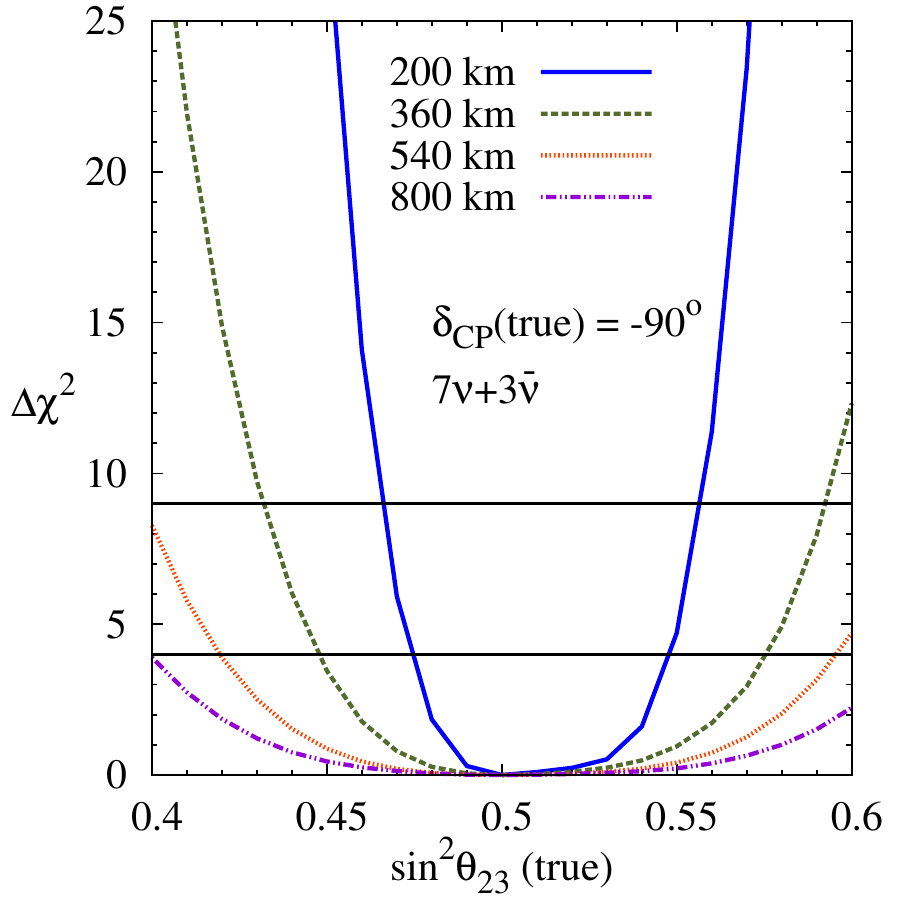}
\includegraphics[width=0.49\textwidth]
{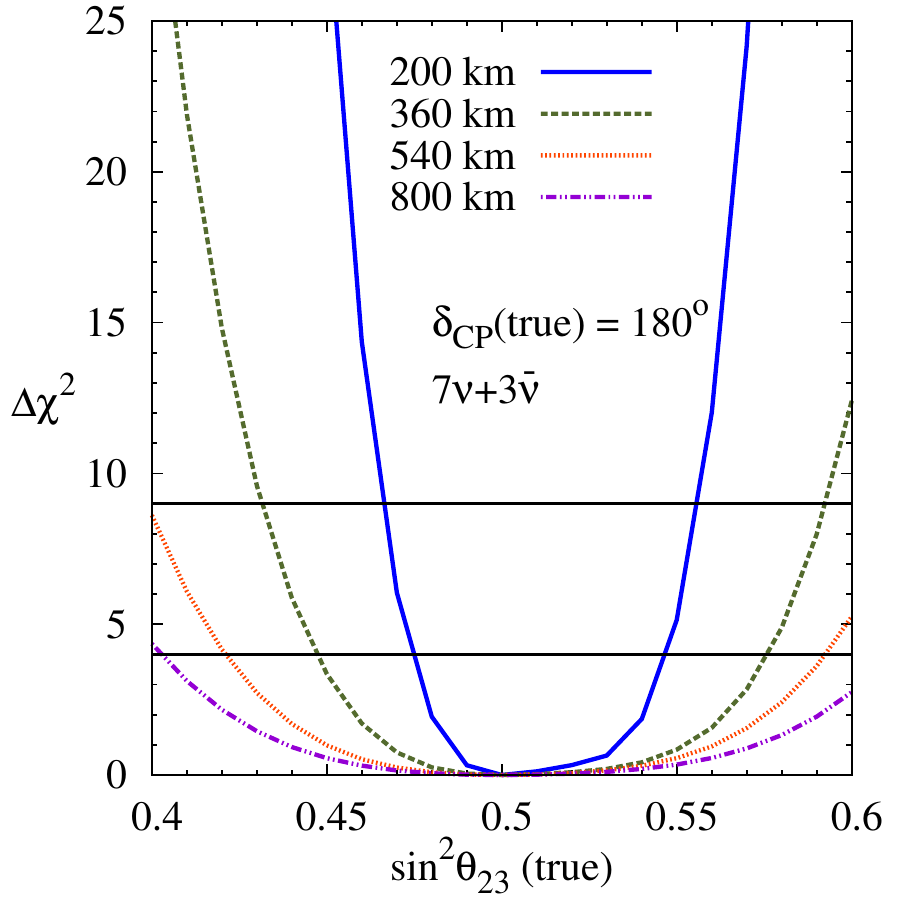}
\caption{\footnotesize{$\dxxmin$ for a non-maximal $\tz$ discovery
vs. $\sin^2\tz$(true) for the ESS$\nu$SB set-up. NH has been 
assumed to be the true hierarchy and the choice of 
run-plan has been taken to be $7\nu+3\anu$ years of running. Results corresponding 
to various choices: 200 km, 360 km, 540 km and 800 km 
for the baseline have been shown. The top-left/top-right/bottom-left/bottom-right
panel corresponds to $0$/$90^\circ$/$-90^\circ$/$180^\circ$ assumed as $\dcp$(true). 
The horizontal black lines show 2\sig and 3\sig confidence level values.}}
\label{non-maximal_baselines}
\end{figure}

As discussed in the Introduction, currently different data sets have a conflict 
regarding the best-fit value of $\theta_{23}$ and its deviation from 
maximal mixing. While global analysis of all data hint at best-fit 
$\theta_{23}$ being non-maximal, these inferences depend on the 
assumed true mass hierarchy and are also not statistically very 
significant. 
Therefore, these results would need further corroboration in the next-generation experiments. 
If the deviation of $\theta_{23}$ from maximal mixing is indeed small,
it may be difficult for the present generation experiment to 
establish a deviation from maximality. It has been checked that the 
combined results from T2K and NO$\nu$A will be able to distinguish 
a non-maximal value of $\theta_{23}$ from the maximal value $\pi/4$ 
at $3\sigma$ C.L. if $\sin^2\theta_{23}$(true)$ \lesssim 0.45$ and $\gtrsim 0.57$. 
In such a situation, it
will be interesting to know how well the ESS$\nu$SB set-up can establish
a non-maximal $\sin^2\tz$. In Fig. \ref{non-maximal_baselines}, we
show the sensitivity of various baselines towards establishing a 
non-maximal $\sin^2\tz$. These plots show the $\dxx$ as a 
function of the true $\sin^2\tz$, where $\dxx$ is as defined in 
section \ref{simulations}.

The results are shown for the prospective baselines of 200 km,
360 km, 540 km and 800 km. The top left (right) panel corresponds 
to the choice of $\dcp$ (true) of $0$ ($90^\circ$). The bottom
left (right) panel corresponds to the choice of true $\dcp$
of $-90^\circ$ ($180^\circ$). The true hierarchy for all these 
plots is assumed to be NH and the run-plan is taken to be 
$7\nu+3\anu$. Here, we have marginalized the $\dxx$ over the 
hierarchy. It can be seen from Fig. \ref{non-maximal_baselines}
that the best sensitivity occurs for the 200 km baseline.
For true $\dcp=0$, a 3\sig determination of non-maximal $\sin^22\tz$
can be made if $\sin^2\tz\lesssim0.47$ or if $\sin^2\tz\gtrsim0.56$.
A 5\sig determination is possible if $\sin^2\tz\lesssim0.45$ 
or if $\sin^2\tz\gtrsim0.57$. We checked that the contribution to
the sensitivity from the appearance channels is small compared to
that from the disappearance channels. 
This is reflected in the 
fact that there is a small dependence of $\dxx$ on the assumed
true value of $\dcp$. 
An interesting observation is that the 
$\dxx$ curve is not symmetric around the $\sin^2\tz=0.5$ line.
It seems that, as far as observing a deviation from maximality is 
concerned, the lower octant is more favored than the higher octant.
The reason behind this feature is the following.
The sensitivity here, is mostly governed by the disappearance data
in which the measured quantity is $\sin^22\theta_{\mu\mu}$. Since
$\sin^2\tz=\sin^2\theta_{\mu\mu}/\cos^2\ty$ 
\cite{Nunokawa:2005nx,deGouvea:2005hk,Raut:2012dm}, the $\ty$ correction
shifts the $\theta_{23}$ values towards $45^\circ$ in the lower
octant and away from $45^\circ$ in the higher octant. This results in the 
shifting of the curve towards the right in $\sin^2\theta_{23}$ and is reflected
as the asymmetric nature of the curve. 
We have checked that for the (now) academic case of $\theta_{13}$(true)$=0$, 
the $\dxx$ curve is symmetric
around $45^\circ$.

\begin{figure}[h]
\centering
\includegraphics[width=0.49\textwidth]
{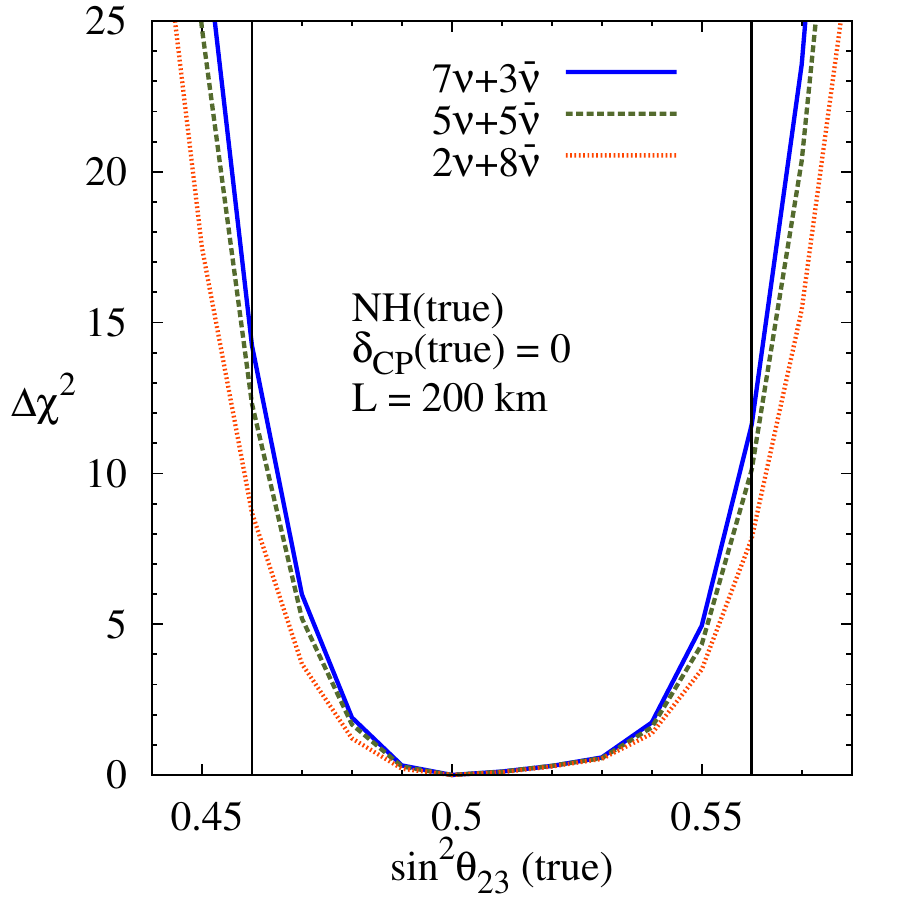}
\caption{\footnotesize{$\dxxmin$ for a non-maximal $\tz$ discovery
vs. $\sin^2\tz$ (true) for the ESS$\nu$SB set-up. NH has been 
assumed to be the true hierarchy and $\dcp$(true)
has been assumed to be 0. The choice of baseline
has been taken to be 200 km. Results corresponding 
to various choices: $2\nu+8\anu$, $5\nu+5\anu$ and $7\nu+3\anu$
years of running for the run-plan have been shown.
The purpose of having vertical lines at $\sin^2\tz~\textrm{true}=0.46$ and 0.56
is to show the effect of run-plan on sensitivity (see Table \ref{tab:maxrunplan} for
discussion on this).}}
\label{non-maximal_runnsys}
\end{figure}

To find an optimal run-plan in the case of deviation from maximality, we 
generated the results for 200 km baseline for ESS$\nu$SB set-up, assuming NH and
$\dcp=0$. Three run-plans were assumed as before: $2\nu+8\anu$, $5\nu+5\anu$
and $7\nu+3\anu$. It can be seen from Fig. \ref{non-maximal_runnsys} that
the best results are observed for $7\nu+3\anu$. Thus, this run-plan seems to
be optimally suited for measurement of deviation from maximality as well.
Note that apparently it seems from Fig. \ref{non-maximal_runnsys} that the
sensitivity in the case of different run-plans are roughly the same despite 
there being huge change of statistics in terms of neutrino and anti-neutrino
data. However, a closer look will reveal that the $\dxx$ indeed changes as expected 
with the increase in the total statistics collected by the 
experiment and in fact it is the very sharp rise of the curves which 
hides the difference. To illustrate 
this further, we show in Table \ref{tab:maxrunplan} the $\dxx$ values corresponding 
to different run-plans at different true $\sin^2\tz$ values and for two choices of 
$\sin^2\theta_{23}$(true).

\begin{table}[h]
\centering
{\footnotesize
\begin{tabular}{|c|c|c|c|}
 \hline
$\sin^2\tz$ (true)
&$2\nu+8\anu$
&$5\nu+5\anu$
&$7\nu+3\anu$
\\
\hline
0.46&8.7&12.3&14.2\\
0.56&7.8&10.1&11.6\\
\hline
\end{tabular}
}
\caption{\footnotesize{$\dxxmin$ for $\sin^2\tz$ (true)
= 0.46 and 0.56. Here, the sensitivity of the ESS$\nu$SB set-up to the deviation 
from a maximal $\tz$ has been considered. NH has been 
assumed to be the true hierarchy and $\dcp$(true)
has been assumed to be 0. The choice of baseline
has been taken to be 200 km. Results corresponding 
to various choices: $2\nu+8\anu$, $5\nu+5\anu$ and $7\nu+3\anu$
years of running for the run-plan have been shown in different columns.}}
\label{tab:maxrunplan} 
\end{table}

\subsection{Octant resolution}

\begin{figure}[h]
\centering
\includegraphics[width=0.3\textwidth]
{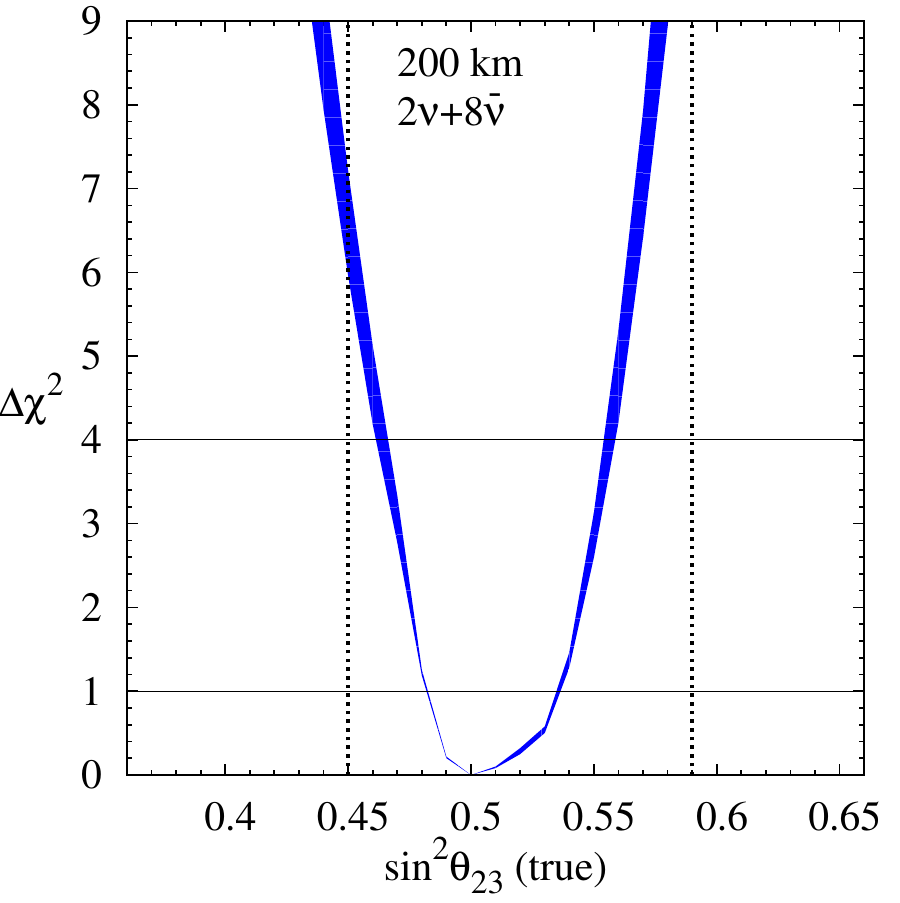}
\includegraphics[width=0.3\textwidth]
{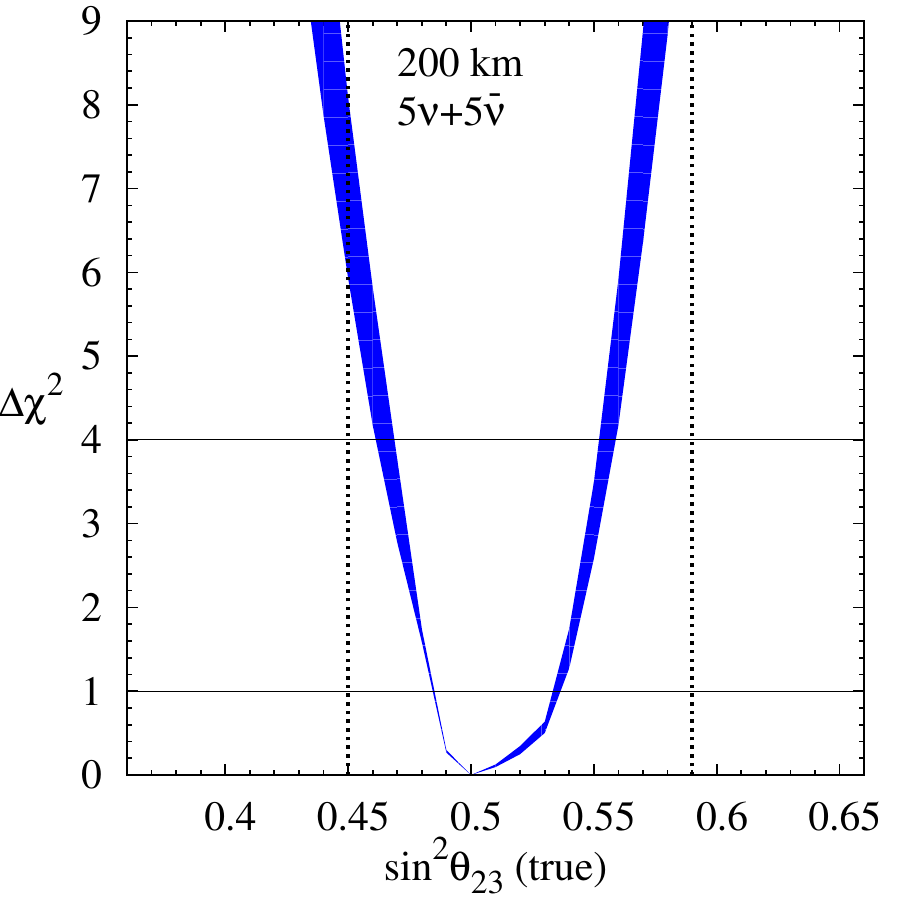}
\includegraphics[width=0.3\textwidth]
{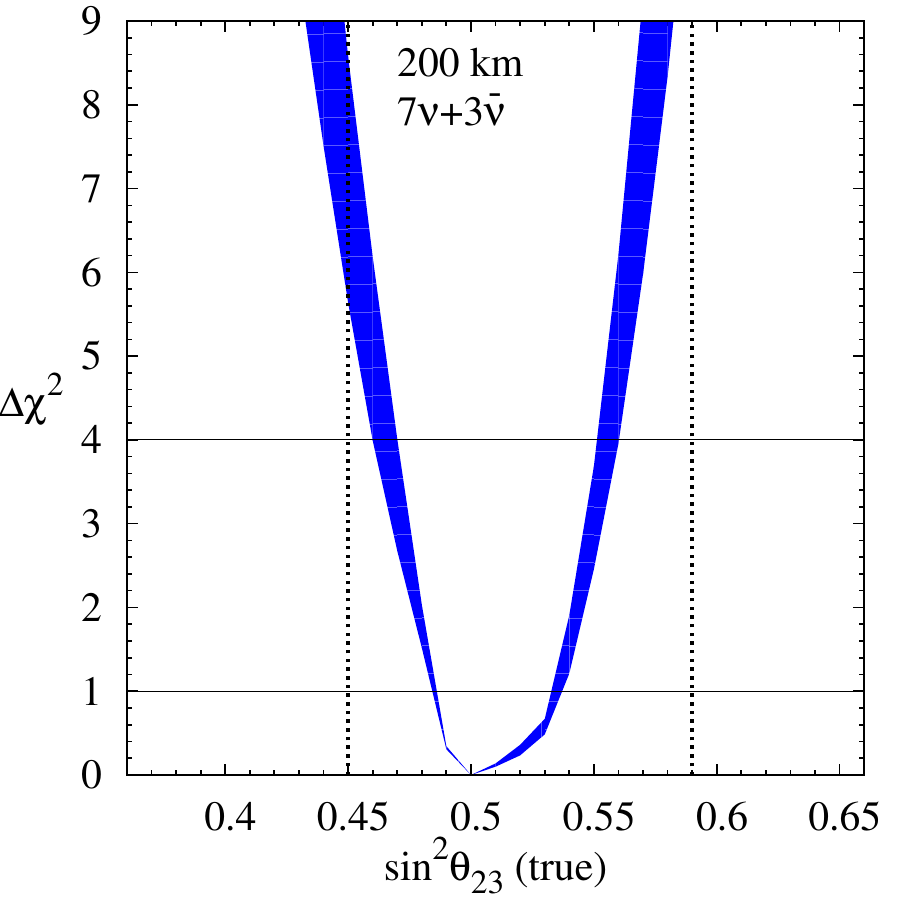}
\includegraphics[width=0.3\textwidth]
{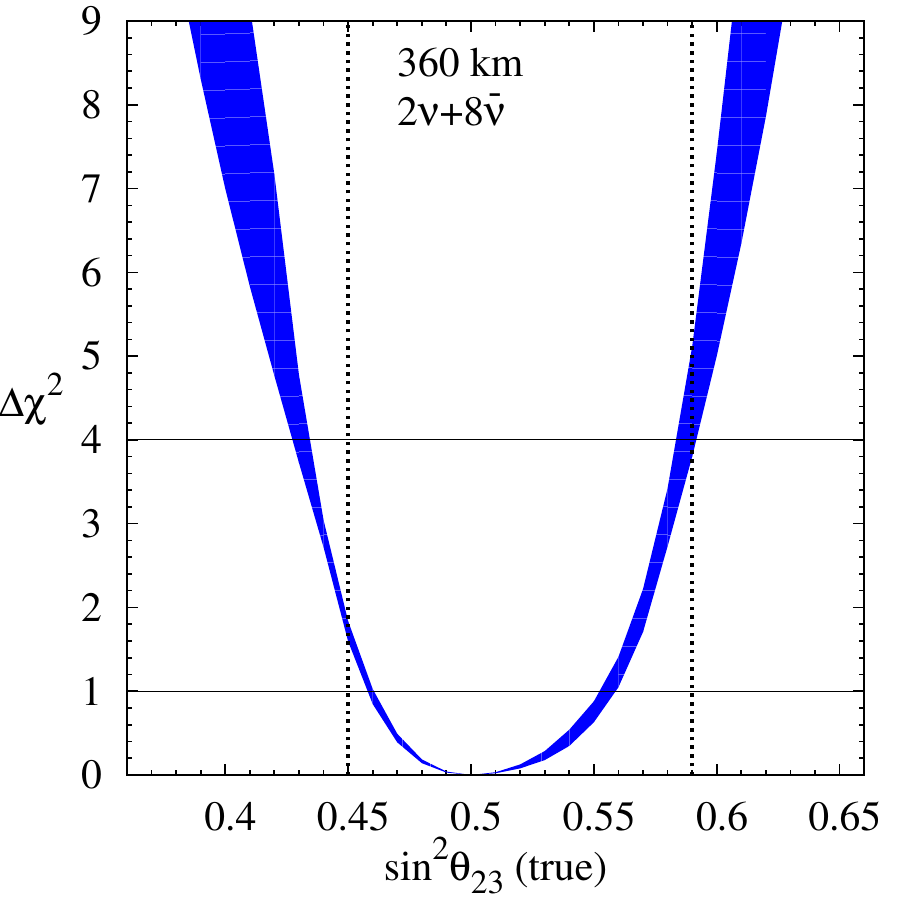}
\includegraphics[width=0.3\textwidth]
{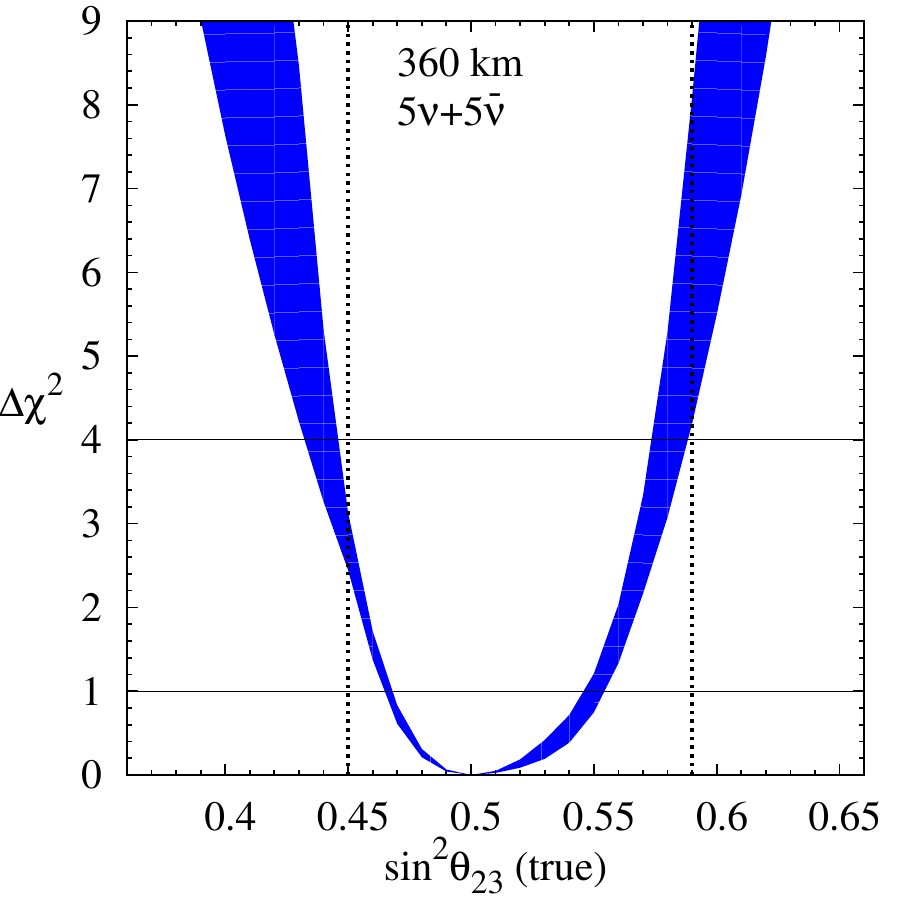}
\includegraphics[width=0.3\textwidth]
{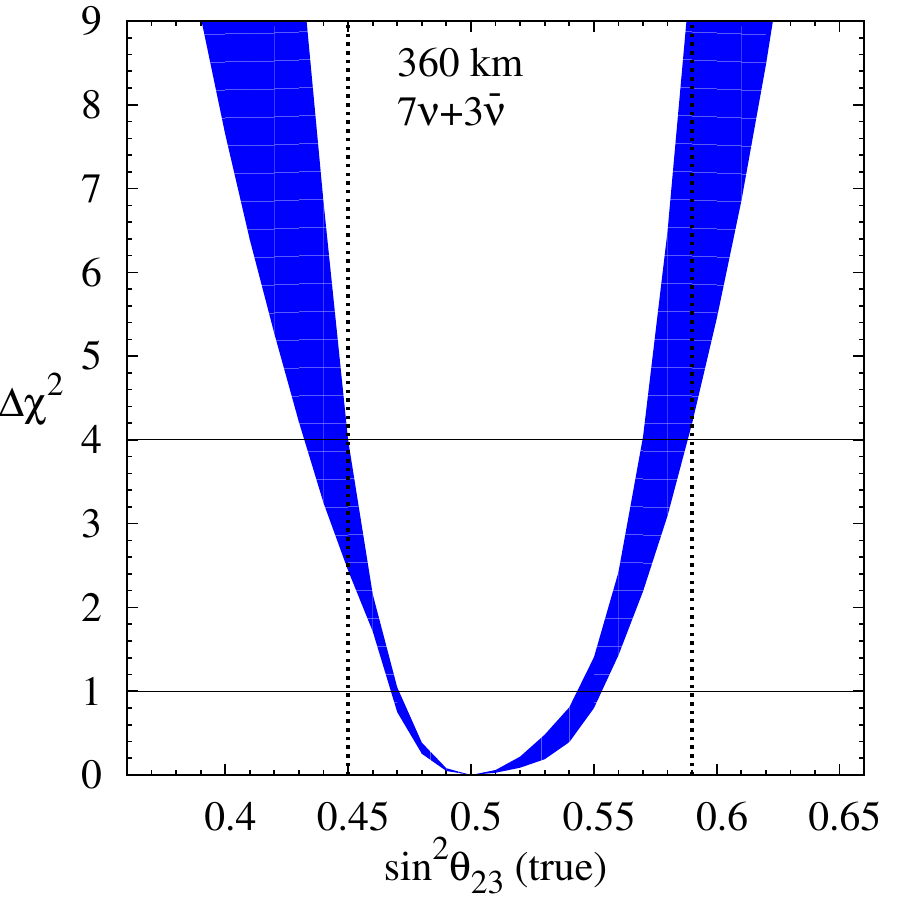}
\includegraphics[width=0.3\textwidth]
{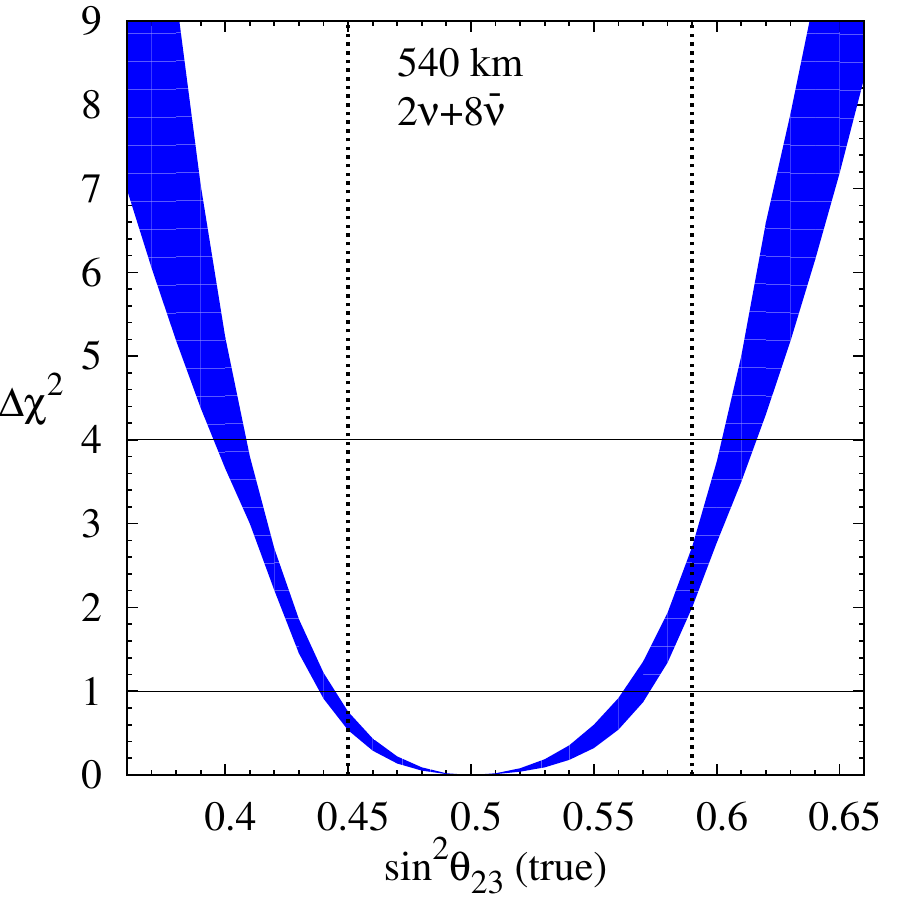}
\includegraphics[width=0.3\textwidth]
{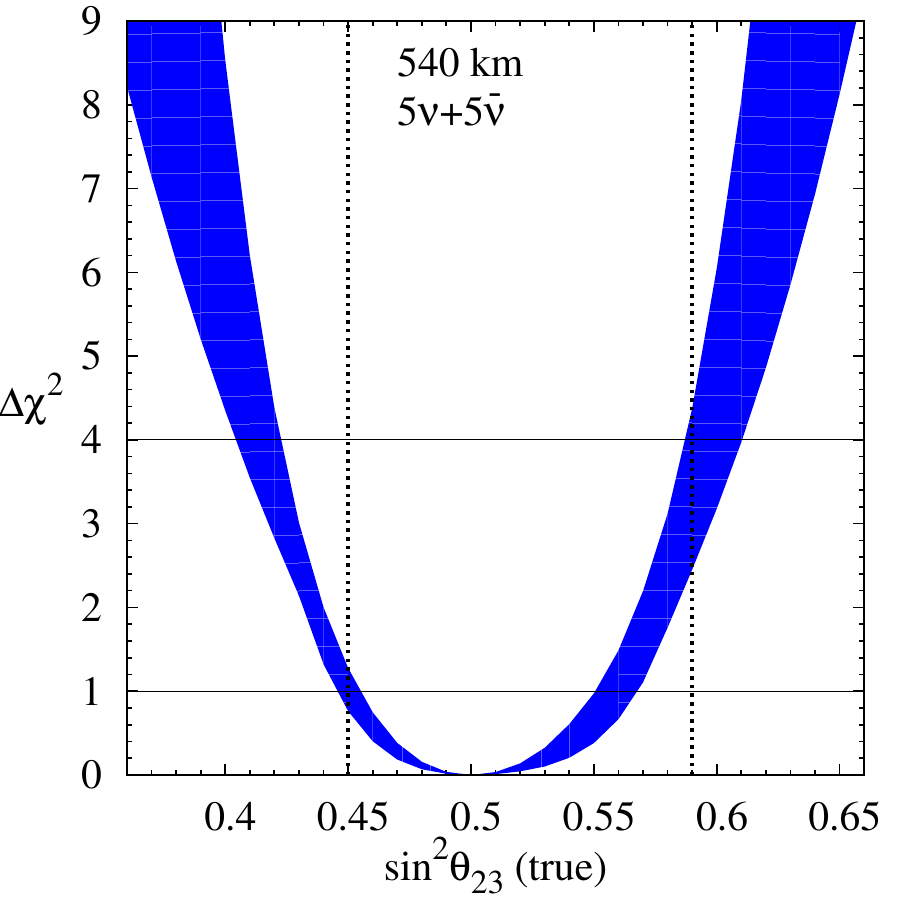}
\includegraphics[width=0.3\textwidth]
{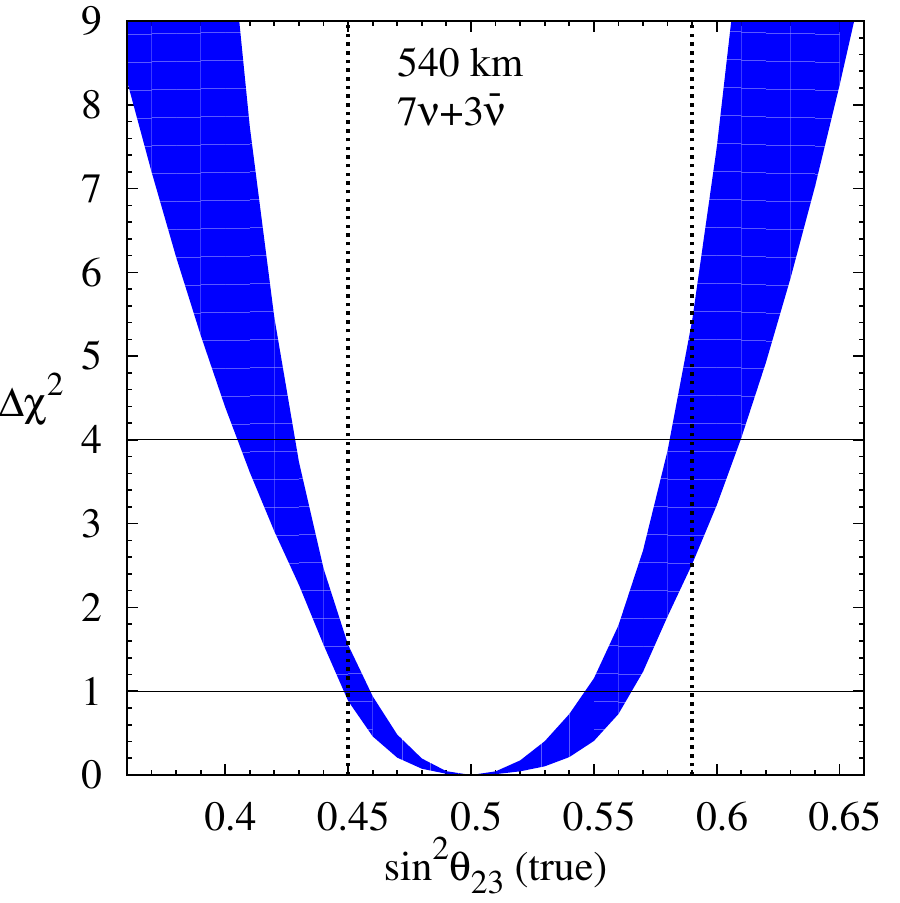}
\includegraphics[width=0.3\textwidth]
{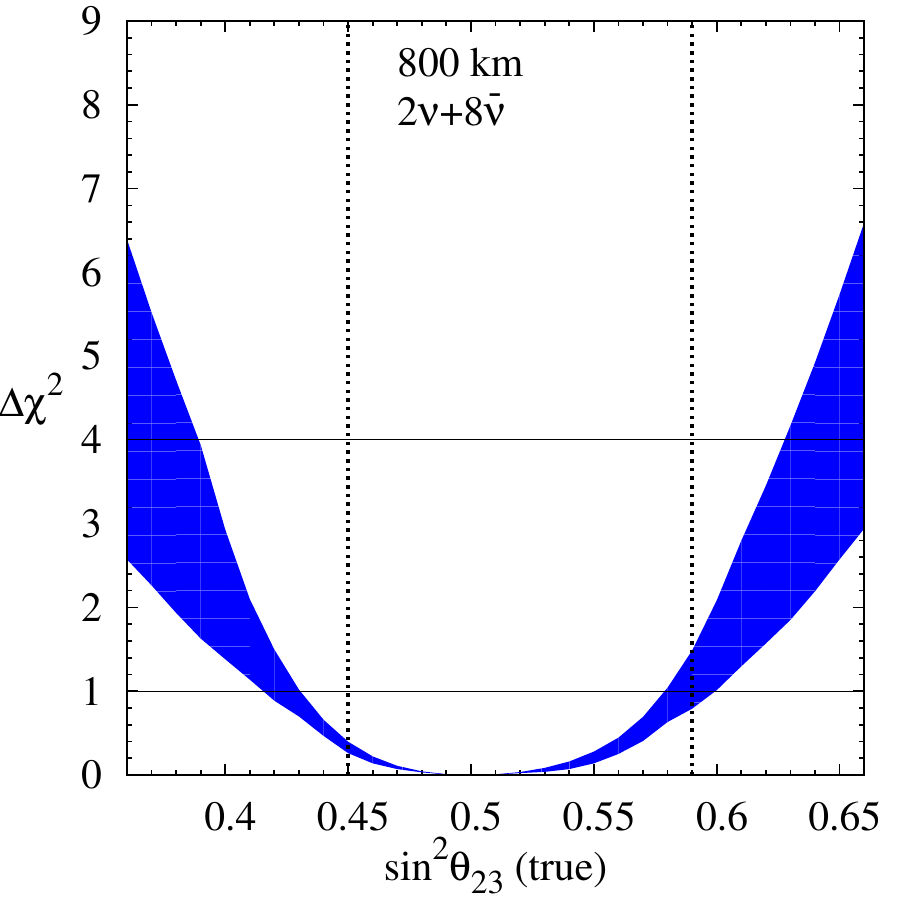}
\includegraphics[width=0.3\textwidth]
{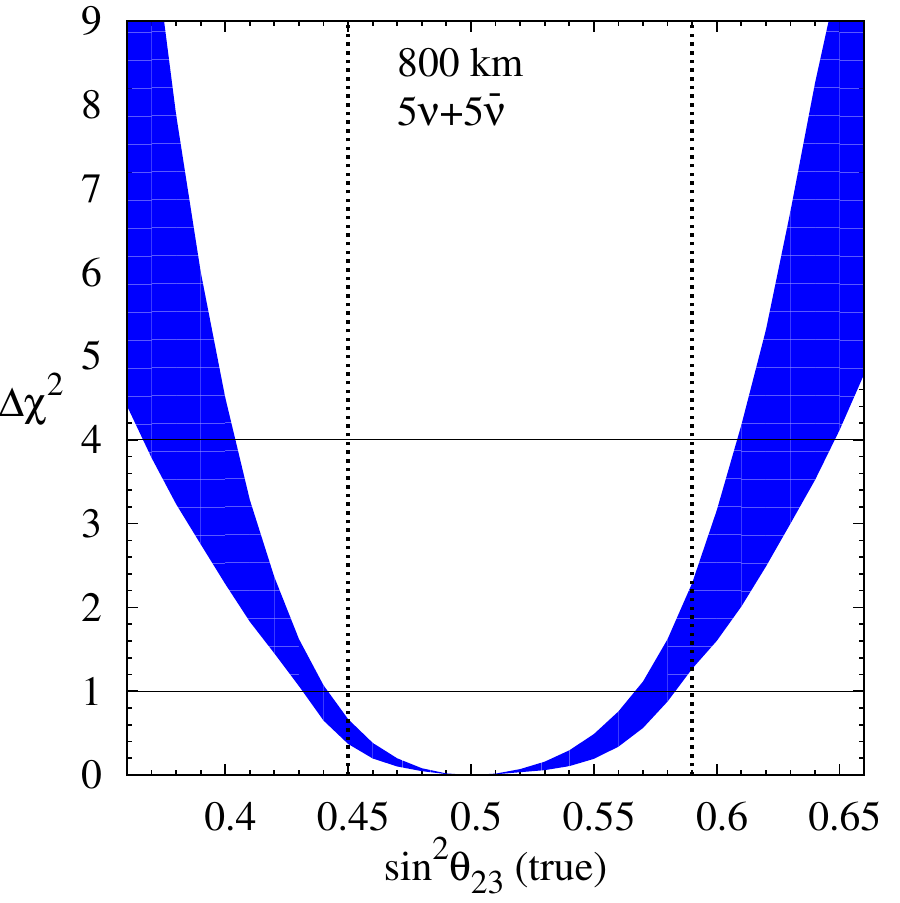}
\includegraphics[width=0.3\textwidth]
{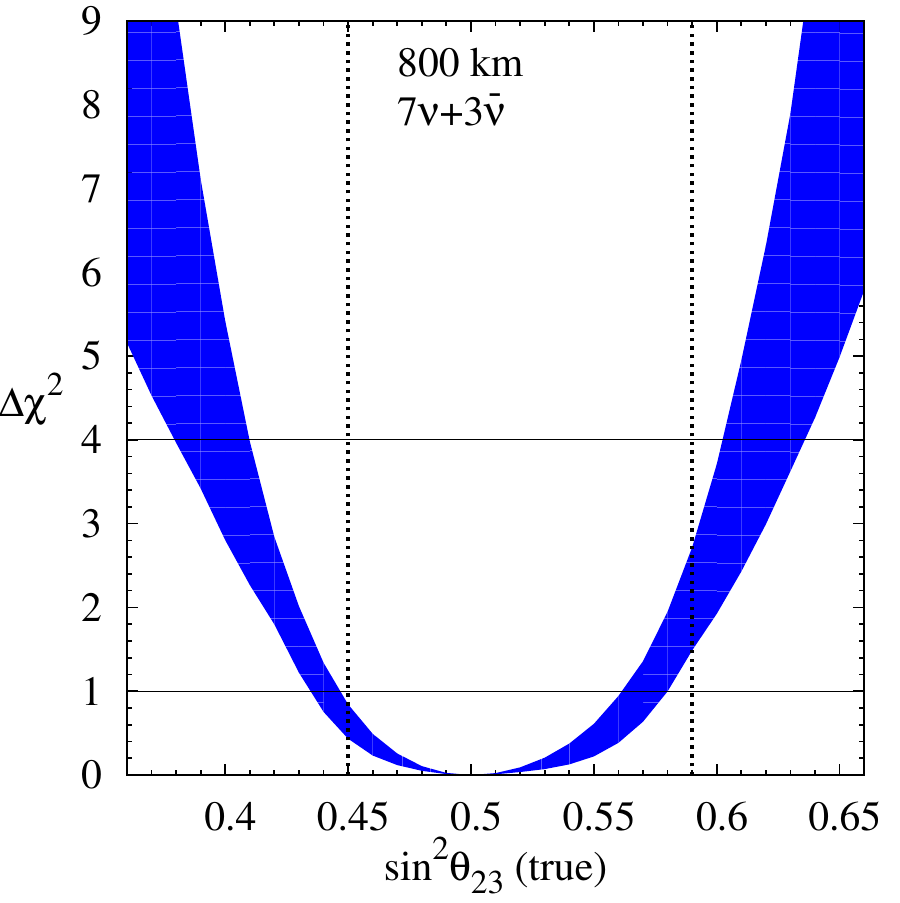}
\caption{\footnotesize{Octant resolution potential as a function 
of $\sin^2\tz$(true) for the ESS$\nu$SB set-up. 
NH has been assumed as the true hierarchy. The variation
in the assumed value of $\dcp$(true) leads to the formation of the 
band. Results corresponding to various run-plans and the assumed
baseline for ESS$\nu$SB set-up have been shown. The rows correspond to
200 km, 360 km, 540 km, and 800 km from top to bottom and the columns
correspond to $2\nu+8\anu$, $5\nu+5\anu$ and $7\nu+3\anu$ 
years of running, from left to right. The horizontal black lines show 1\sig and 2\sig 
confidence level values.}}
\label{octant_runplan_NH}
\end{figure}

\begin{figure}[h]
\centering
\includegraphics[width=0.3\textwidth]
{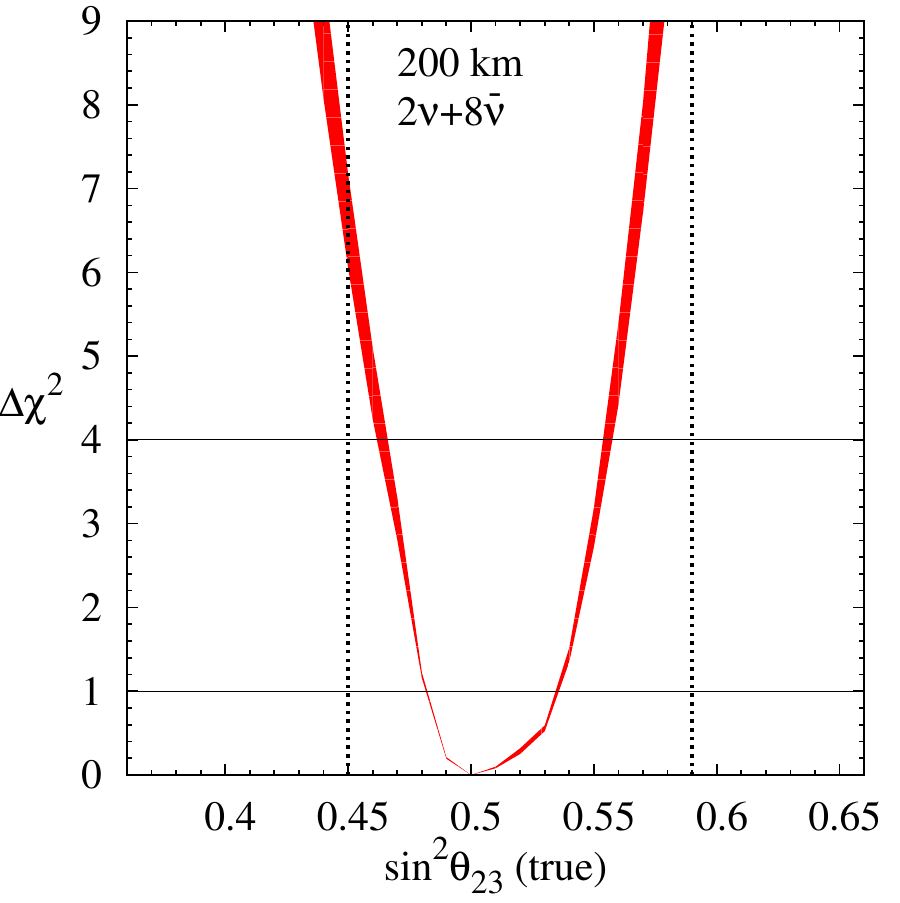}
\includegraphics[width=0.3\textwidth]
{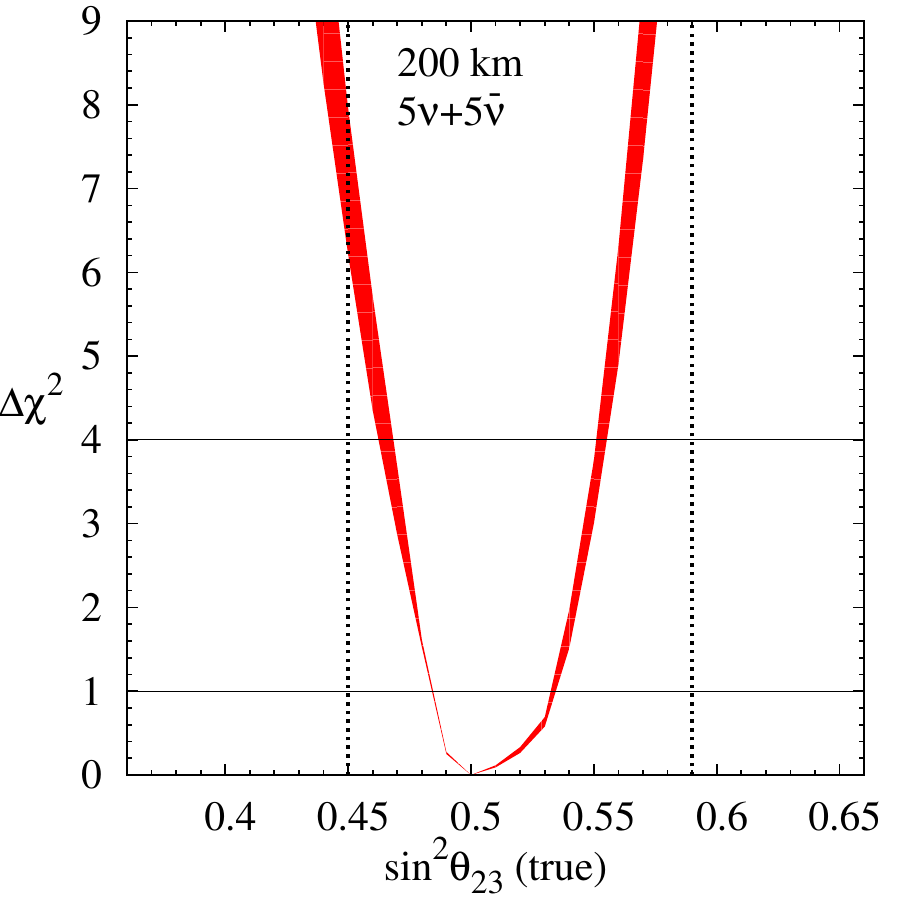}
\includegraphics[width=0.3\textwidth]
{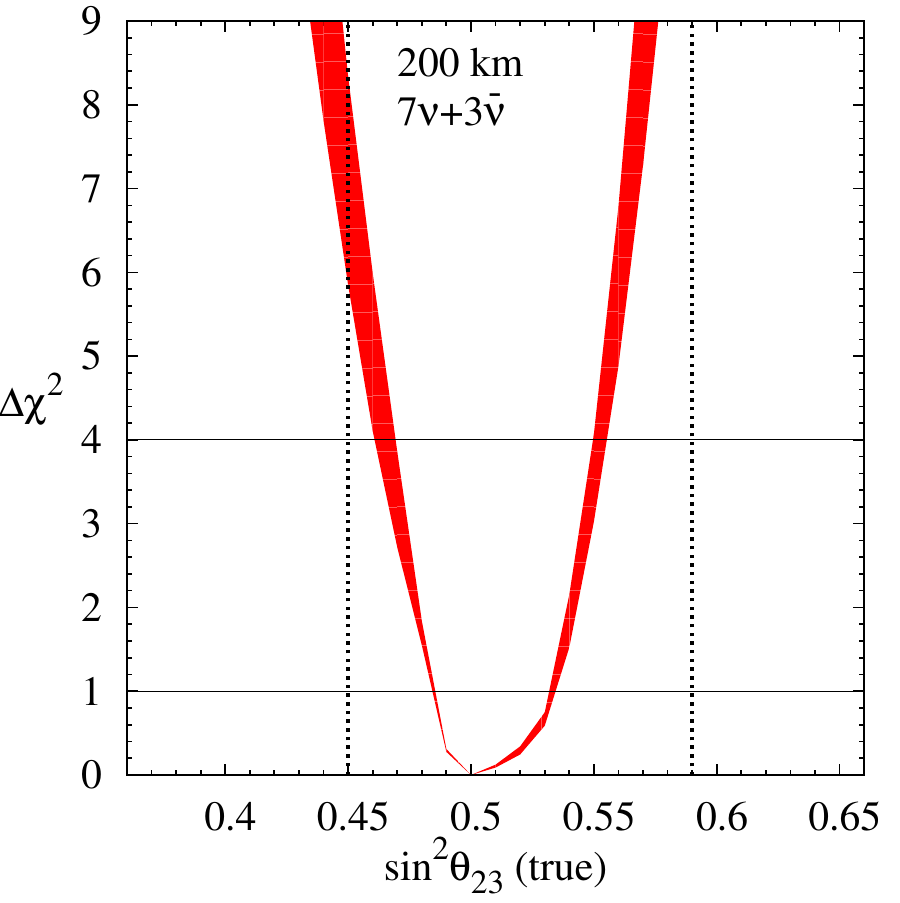}
\includegraphics[width=0.3\textwidth]
{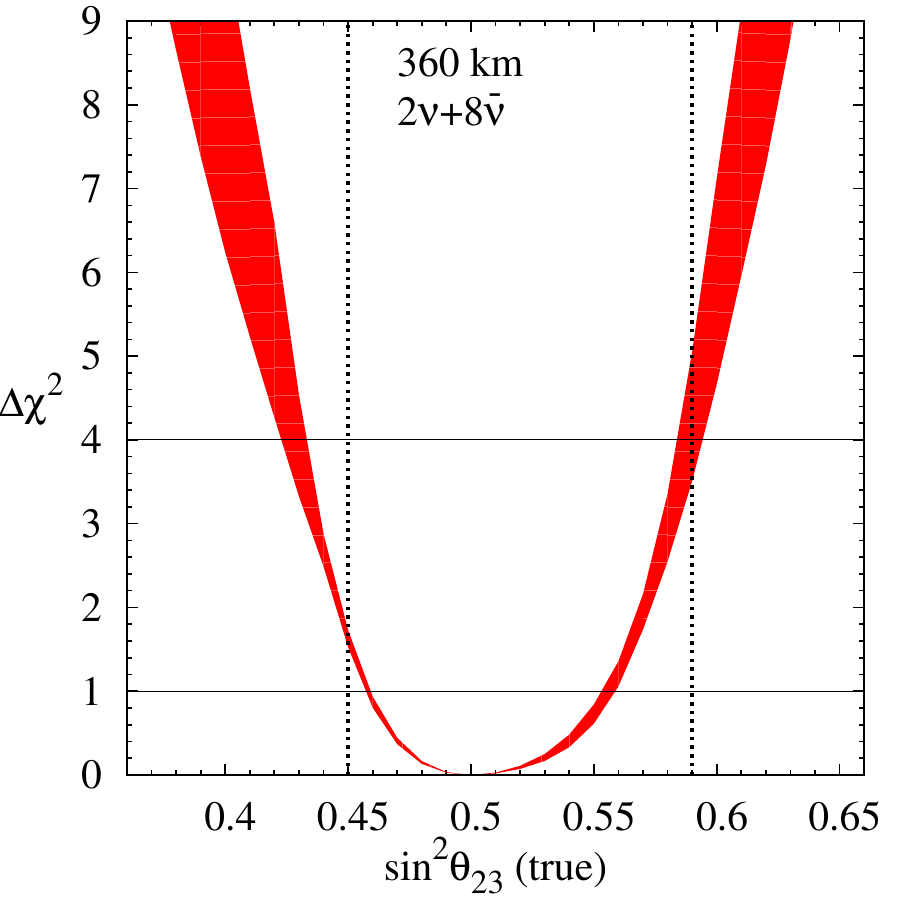}
\includegraphics[width=0.3\textwidth]
{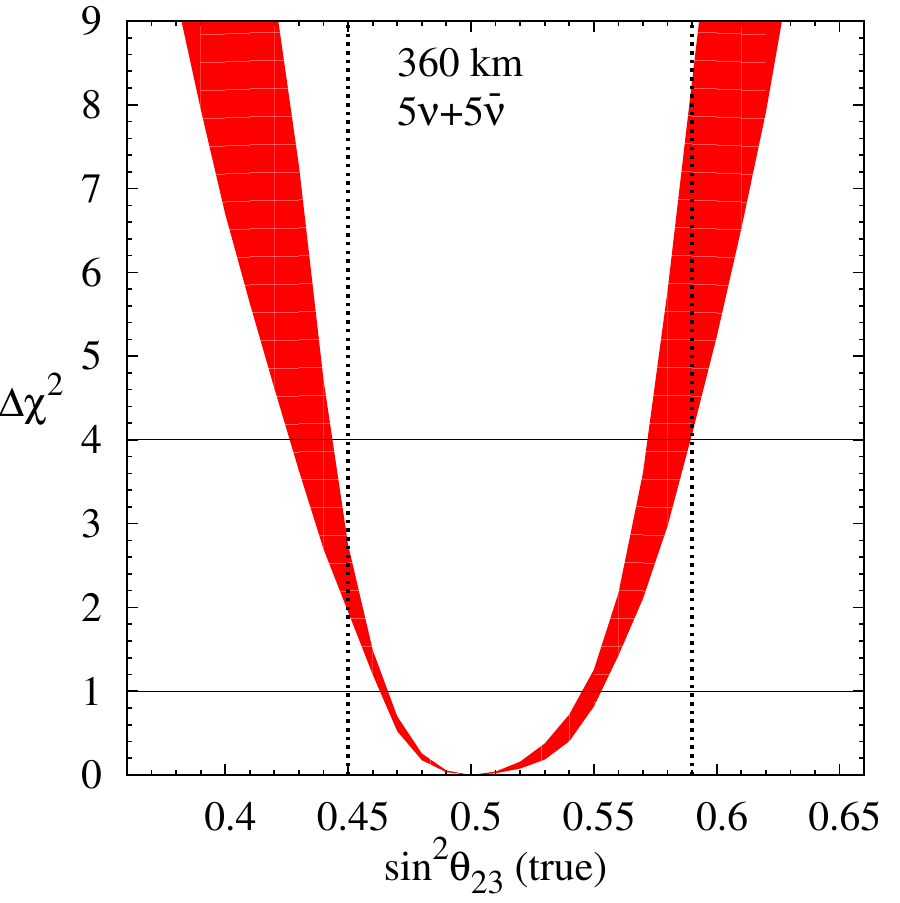}
\includegraphics[width=0.3\textwidth]
{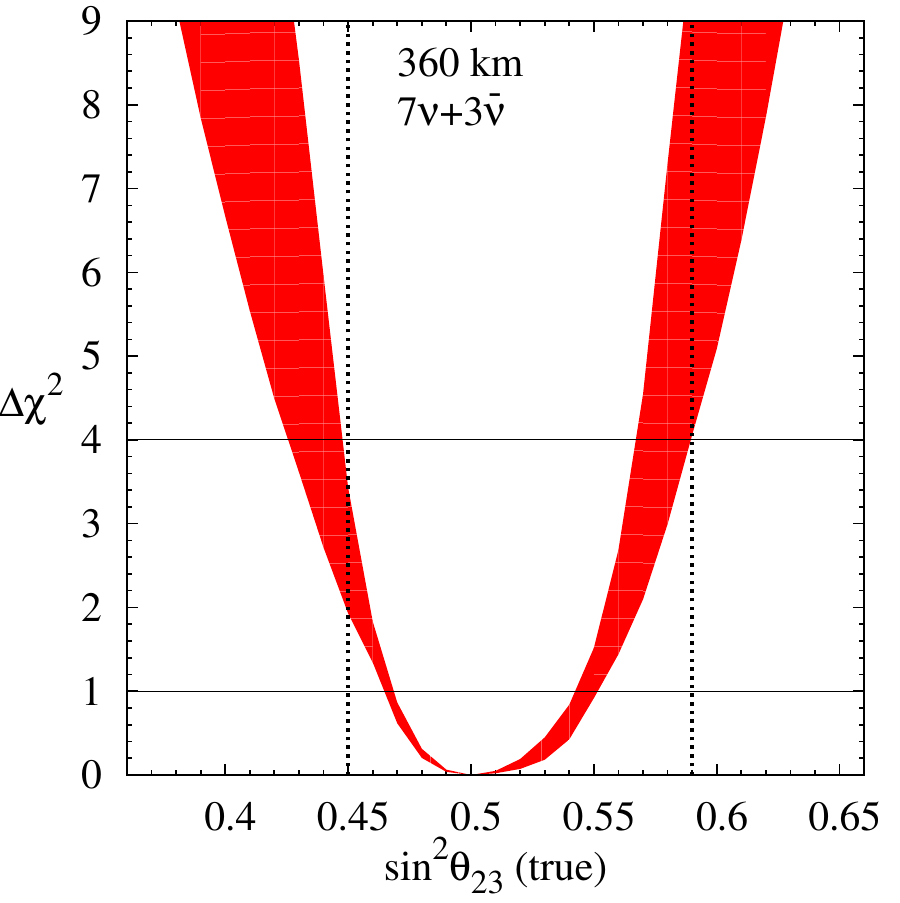}
\includegraphics[width=0.3\textwidth]
{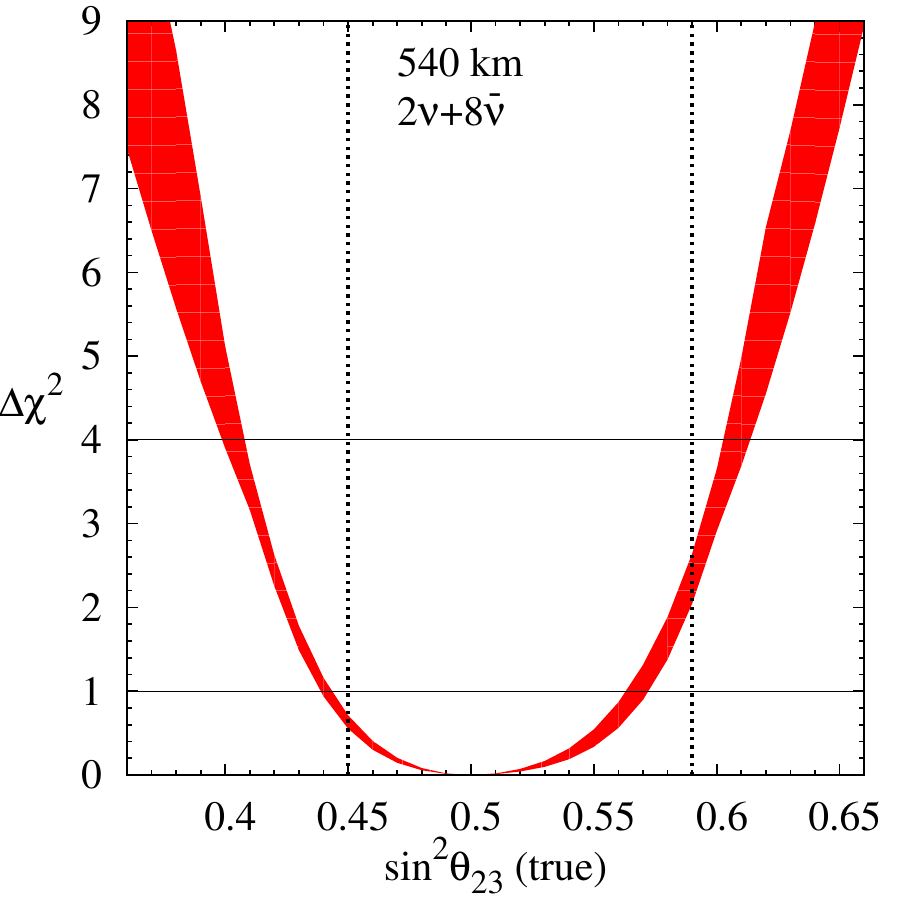}
\includegraphics[width=0.3\textwidth]
{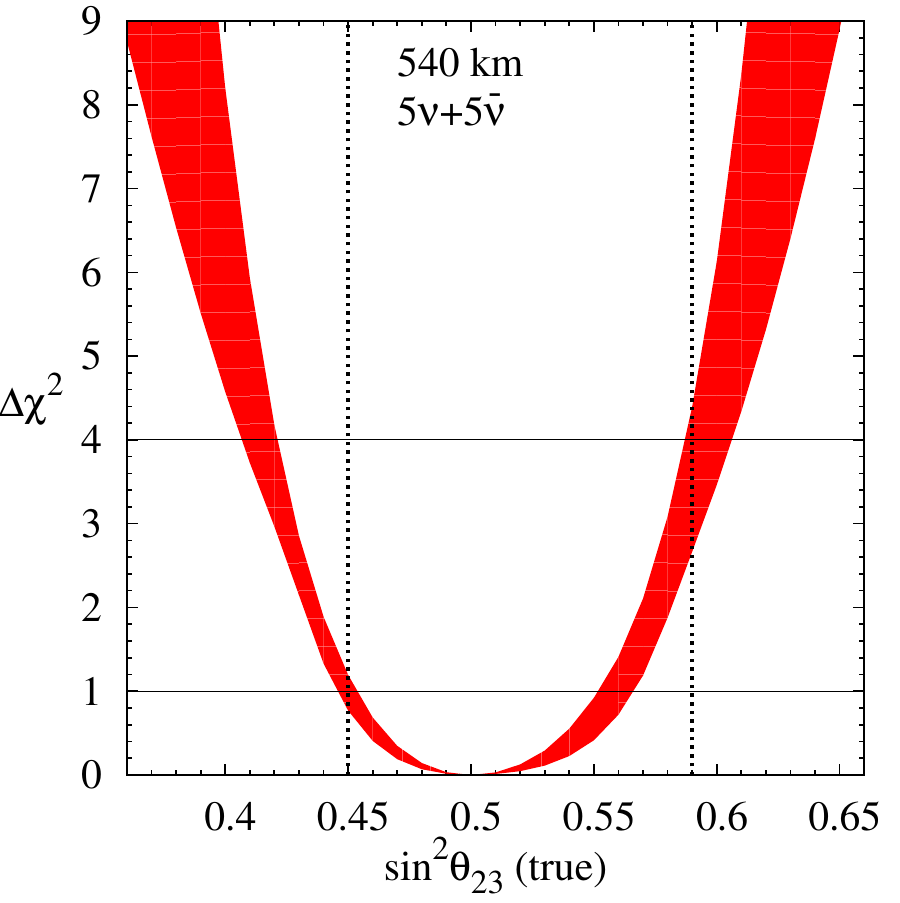}
\includegraphics[width=0.3\textwidth]
{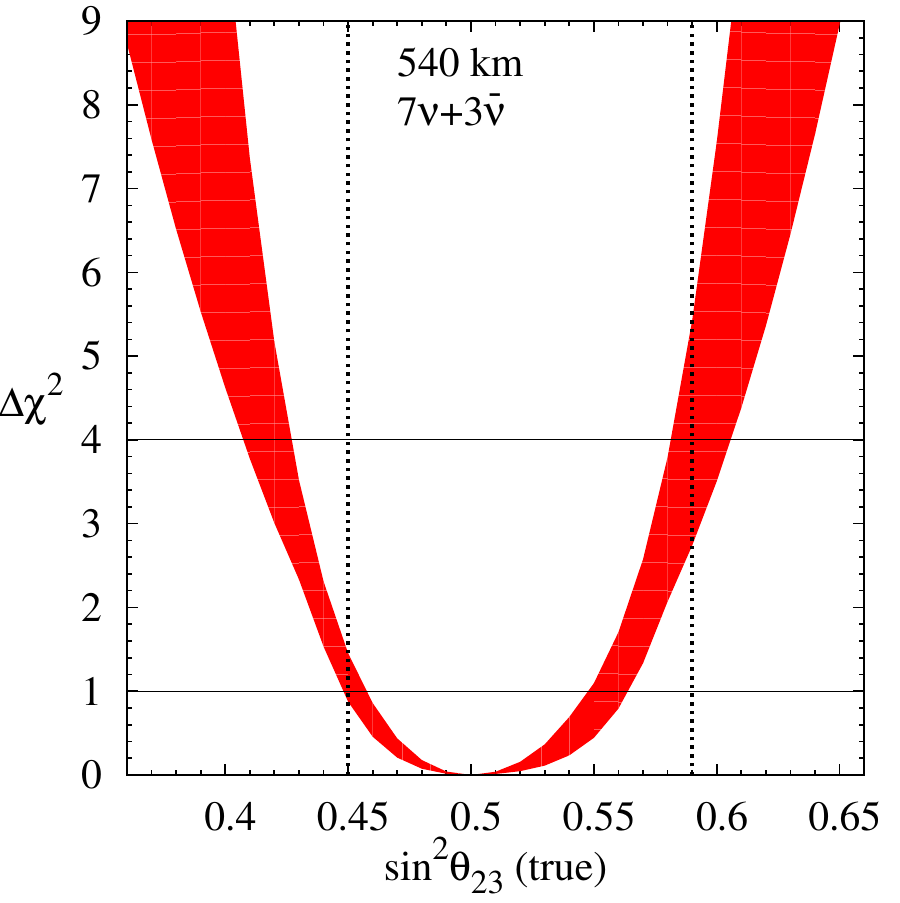}
\includegraphics[width=0.3\textwidth]
{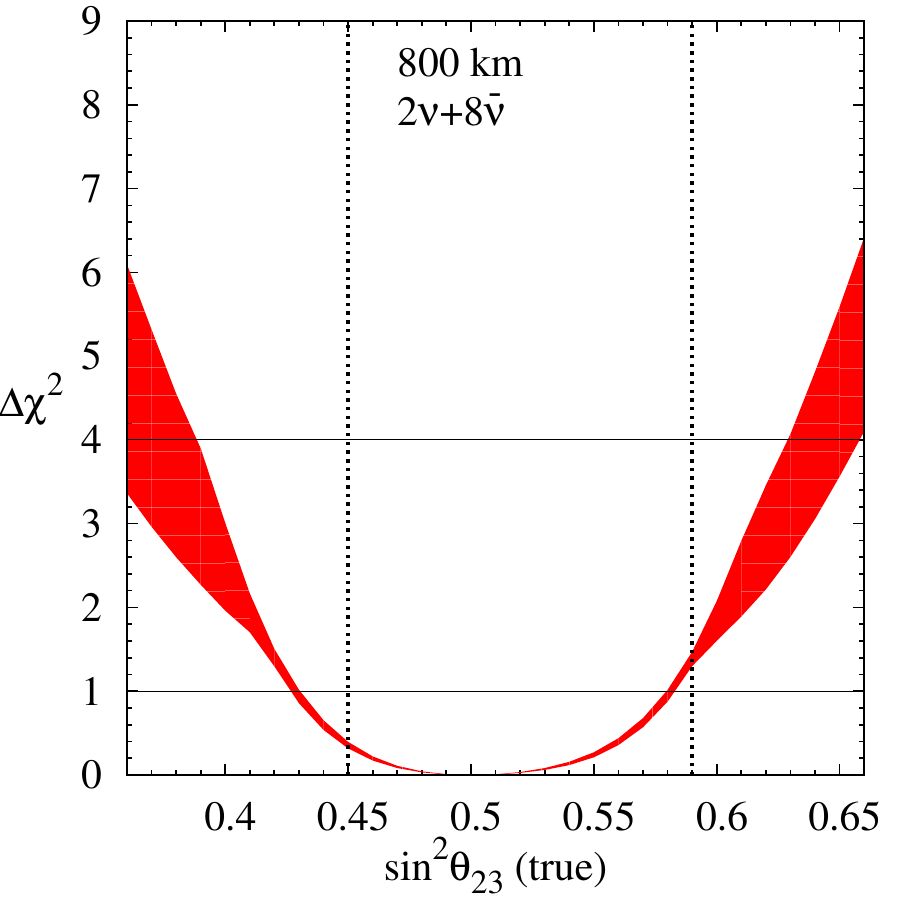}
\includegraphics[width=0.3\textwidth]
{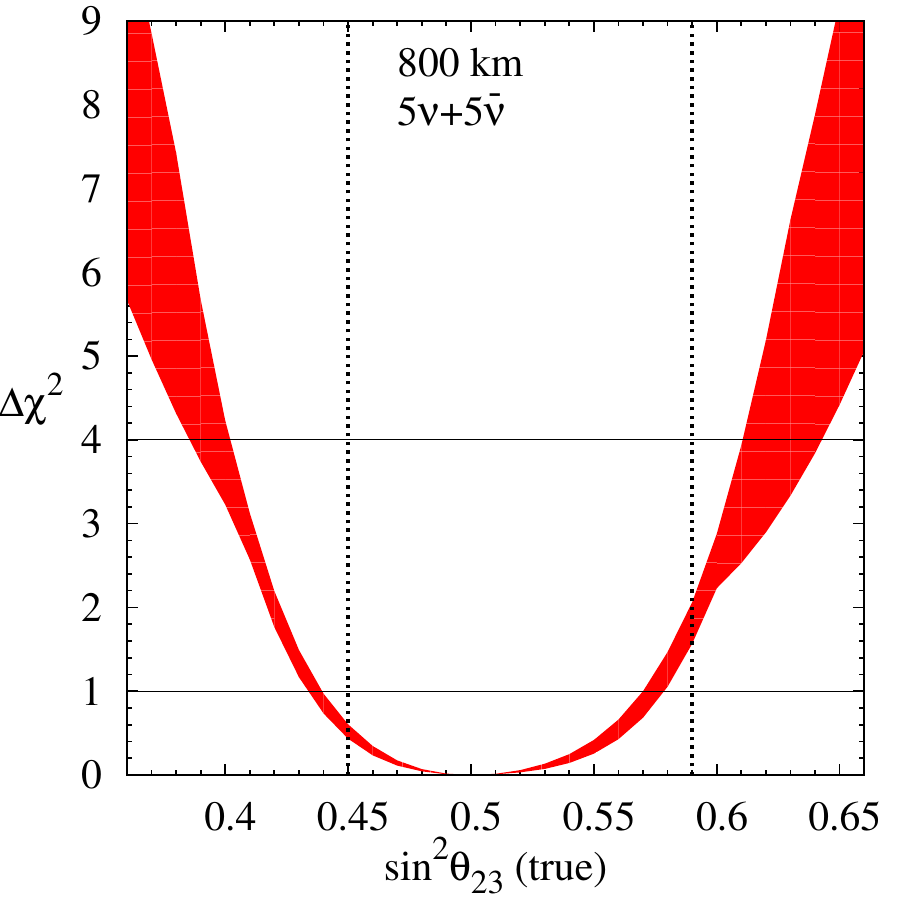}
\includegraphics[width=0.3\textwidth]
{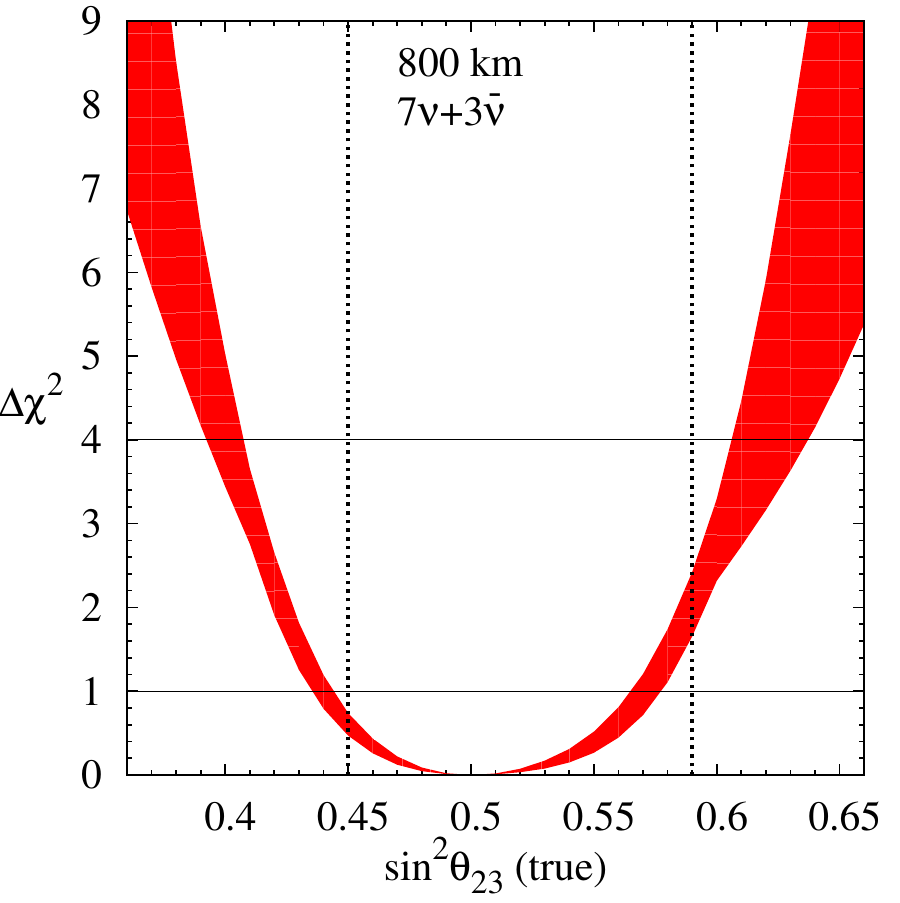}
\caption{\footnotesize{Octant resolution potential as a function 
of $\sin^2\tz$(true) for the ESS$\nu$SB set-up. 
IH has been assumed as the true hierarchy. The variation
in the assumed value of $\dcp$(true) leads to the formation of the 
band. Results corresponding to various run-plans and the assumed
baseline for ESS$\nu$SB set-up have been shown. The rows correspond to
200 km, 360 km, 540 km, and 800 km from top to bottom and the columns
correspond to $2\nu+8\anu$, $5\nu+5\anu$ and $7\nu+3\anu$ 
years of running, from left to right. The horizontal black lines show 1\sig and 2\sig 
confidence level values.}}
\label{octant_runplan_IH}
\end{figure}

In this section, we explore the octant resolving capability
of the ESS$\nu$SB set-up. As discussed in the previous section, we 
generate true event rates at certain $\sin^2\theta_{23}$(true) and 
fit this by marginalizing over the entire $\sin^2\theta_{23}$ range 
in the wrong octant. The $\dxx$ is also marginalized over $|\Delta m^2_{31}|$, 
$\sin^2\theta_{13}$, $\dcp$ and the neutrino mass hierarchy. 
Fig. \ref{octant_runplan_NH} shows the $\dxx$ obtained as a function of 
$\sin^2\tz$(true) assuming NH to the true hierarchy.
The corresponding results for the IH(true) case is shown in 
Fig. \ref{octant_runplan_IH}. We show the results for
200 km, 360 km, 540 km, and 800 km baselines in the
first, second, third, and fourth rows respectively.
The first column corresponds to the $2\nu+8\anu$
run-plan. The second column corresponds to the $5\nu+5\anu$ run-plan 
while the third column corresponds to the $7\nu+3\anu$ run-plan.
The band in each of these plots correspond 
to variation of $\dcp$(true) in the range $[-180^\circ,180^\circ]$.
Thus, for any $\sin^2\theta_{23}$(true), the top-most and the bottom-most
$\dxx$ values lying in the band shows the maximum and minimum
$\dxx$ possible depending on the true value of $\dcp$.

From the plots in Fig. \ref{octant_runplan_NH} and Fig. \ref{octant_runplan_IH}, 
it can be seen that the best choice for $\theta_{23}$ octant resolution seems 
to be the 200 km baseline. Since amongst the various choices, the 200 km 
baseline is the closest to the source, 
it has the largest statistics for both $\nu$ and $\anu$ samples. This 
is the main reason why the 200 km option returns the best octant 
resolution prospects. We have explicitly checked that if the statistics of the 
the other baseline options were scaled to match the one we get for the 
200 km baseline option, they would give $\theta_{23}$ octant sensitivity 
close to that obtained for the 200 km option. 
We can also see from these figures that the best sensitivity is 
expected for the run-plan of $7\nu+3\anu$. 
Note also that
the $5\nu+5\anu$ run-plan is just marginally worse than the $7\nu+3\anu$
plan. However, these two run-plans are better than $2\nu+8\anu$ run-plan.
This again comes 
because of the fact that this option allows for larger statistics 
while maintaining a balance between the $\nu$ and $\anu$ data, 
which is required to cancel degeneracies for maximum 
octant resolution capability as was shown in \cite{Agarwalla:2013ju}. 
The impact of the run plans are again seen to be larger for the 
larger baselines. 
We can also see that the impact of $\dcp$(true) is larger for 
larger baselines. The $\dcp$ band is narrowest for the 
200 km baseline option, implying that this baseline choice
suffers least uncertainty from unknown $\dcp$(true) for 
octant studies. 

Assuming NH(true) and with the 200 km baseline and $7\nu+3\anu$ run-plan option, 
one can expect to resolve the correct octant of $\theta_{23}$ at the $3\sigma$ level for 
$\sin^2\theta_{23}$(true)$\lesssim 0.43$ and $\gtrsim 0.59$ irrespective of $\dcp$(true). 
Correct octant can be identified with this option at 5\sig confidence level for 
$\sin^2\theta_{23}$(true)$\lesssim 0.37$ and $\gtrsim 0.63$ for all values of $\dcp$(true). 
For IH(true) the corresponding values for $3\sigma$ 
($5\sigma$) sensitivity are 
$\sin^2\theta_{23}$(true)$\lesssim0.43(0.37)$ and $\gtrsim0.59(0.62)$. 
These numbers and a comparison of Fig. \ref{octant_runplan_NH} and \ref{octant_runplan_IH}
reveals that the octant sensitivity of 
the ESS$\nu$SB set-up does not depend much on the 
assumed true mass hierarchy.  
The octant sensitivity for both true hierarchies and all run-plan 
options is seen to deteriorate rapidly with the increase in the 
baseline. For the 540 km baseline option, we find that 
even for $\sin^2\theta_{23}$(true)$>0.35$ and $<0.63$, we 
do not get a $3\sigma$ resolution of the octant for 100\% 
values of $\dcp$(true). 

\begin{figure}[h]
\centering
\includegraphics[width=0.49\textwidth]
{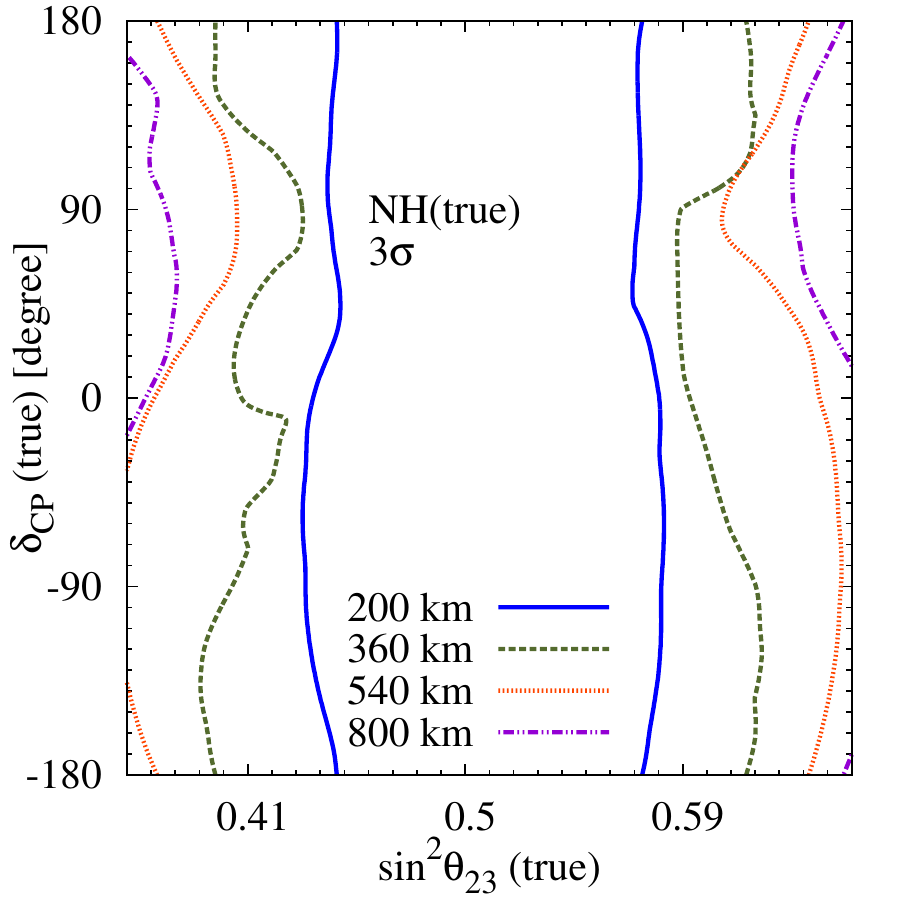}
\includegraphics[width=0.49\textwidth]
{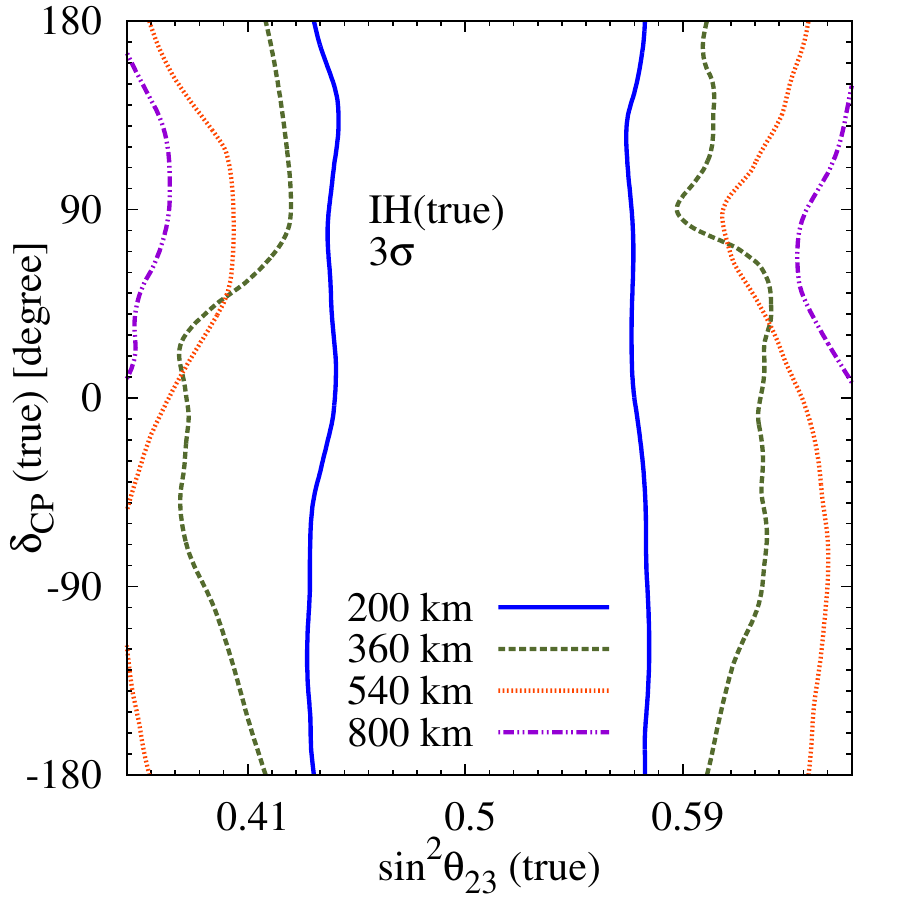}
\caption{\footnotesize{3\sig C.L. contours in the $\sin^2\tz$(true)-$\dcp$(true)
plane for the octant-resolution sensitivity of the ESS$\nu$SB set-up. 
The left(right) panel corresponds to NH(IH) assumed as the true hierarchy.
Results for various possible choices of baseline have been shown.
The run-plan considered here is $7\nu+3\anu$ years of running.}}
\label{octant_contour}
\end{figure}

To show the impact of $\dcp$(true) on the determination 
of the octant of $\theta_{23}$ at ESS$\nu$SB, we show  
in Fig. \ref{octant_contour} the 
$3\sigma$ contours in the $\sin^2\theta_{23}$(true)-$\dcp$(true)
plane for different baselines. We assume the $7\nu+3\anu$ run-plan 
for this figure. The left hand panel shows the contours for NH(true) 
while the right hand panel is for IH(true). The different lines 
show the contours for the different baselines. 
Comparison of the different lines reveals
that the 200 km baseline is better-suited
for the resolution of octant. 
Not only does it gives the best octant determination potential, it also 
shows least 
$\dcp-\sin^2\tz$ correlation.
For other baselines, the contours fluctuate more depending on $\dcp$(true) 
as for these baselines, the ESS fluxes peak close to
the second oscillation maximum, where a larger sensitivity
to $\dcp$ exists. Hence, we see larger dependence of the 
sensitivity on the assumed true value of $\dcp$. In particular, 
the performance is seen to be worst for $\dcp$(true)$\simeq -90^\circ$ 
and best for $90^\circ$. 

\section{Summary and Conclusions} 
\label{sec:conclusions}

The ESS proposal is envisaged as a major European facility for 
neutron source, to be used for both research as well as the 
industry. A possible promising extension of this project 
could be to use it simultaneously to produce a high intensity 
neutrino superbeam to be used for oscillation physics. 
Since the energy of the beam is comparatively lower, 
it has been proposed to do this oscillation experiment at the 
second oscillation maximum, for best sensitivity to 
CP violation discovery. 
In this work we have made a comparative study of all oscillation 
physics searches with ESS$\nu$SB, allowing for all possible 
source-detector distances and with different run-plan 
options for running the experiment in the neutrino and anti-neutrino 
modes. 

In particular, we have evaluated the sensitivities of the ESS$\nu$SB 
proposal towards the discovery of CP violation in the lepton sector,
achievable precision on atmospheric parameters, deviation of
$\sin^2\tz$ from 0.5, and finally the octant in which it lies.
We have considered the prospective baselines - 200 km, 360 km, 
540 km, and 800 km for the resolution of the above mentioned unknowns. 
We also tested different run-plans i.e. varying combination of $\nu$ and $\anu$ data 
with a total of 10 years of running. We considered $2\nu+8\anu$, 
$5\nu+5\anu$ and $7\nu+3\anu$. In the case of CP violation,
we find that the best sensitivity comes for 540 km baseline where
70\% coverage is possible in true $\dcp$ at 3\sig while a 
45\% coverage is possible at 5\sig. For the 200 km baseline,
we find that 60\% coverage is possible at 3\sig and 32\%
coverage is possible at 5\sig. We further find that all the 
three run-plans give the same coverage at 2\sig C.L. but,
at 5\sig C.L., a better coverage is possible with the $7\nu+3\anu$
run-plan. For determination of deviation 
of $\theta_{23}$ from maximality, the best
sensitivity is expected for the 200 km baseline with the 
$7\nu+3\anu$ run-plan, as this combination provides the 
largest statistics. For true $\dcp=0$, a 3\sig determination of non-maximal $\sin^22\tz$
can be made if true value of $\sin^2\tz\lesssim0.47$ or $\gtrsim0.56$.
A 5\sig determination is possible if the true value of 
$\sin^2\tz\lesssim0.45$ or $\gtrsim0.57$. 
In the case of octant also, we find that the 
200 km baseline and $7\nu+3\anu$ run-plan 
provides the best sensitivity. We find that,
assuming NH to be the true hierarchy, a 3\sig resolution of octant is
possible if $\sin^2\theta_{23}$(true) $\lesssim 0.43$ and $\gtrsim 0.59$ 
for all values of $\dcp$(true). A $5\sigma$ determination could 
be possible if $\sin^2\theta_{23}$(true) $\lesssim 0.37$ and $\gtrsim 0.63$. 

\begin{figure}[h]
\centering
\includegraphics[width=0.49\textwidth]
{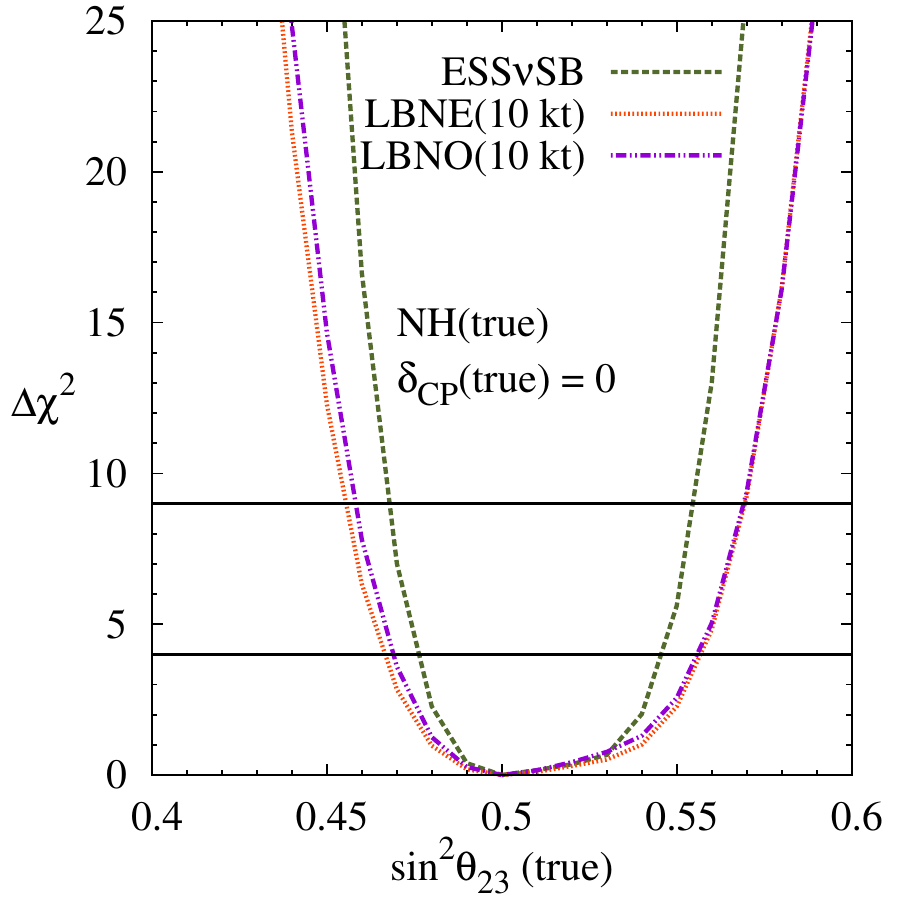}
\includegraphics[width=0.49\textwidth]
{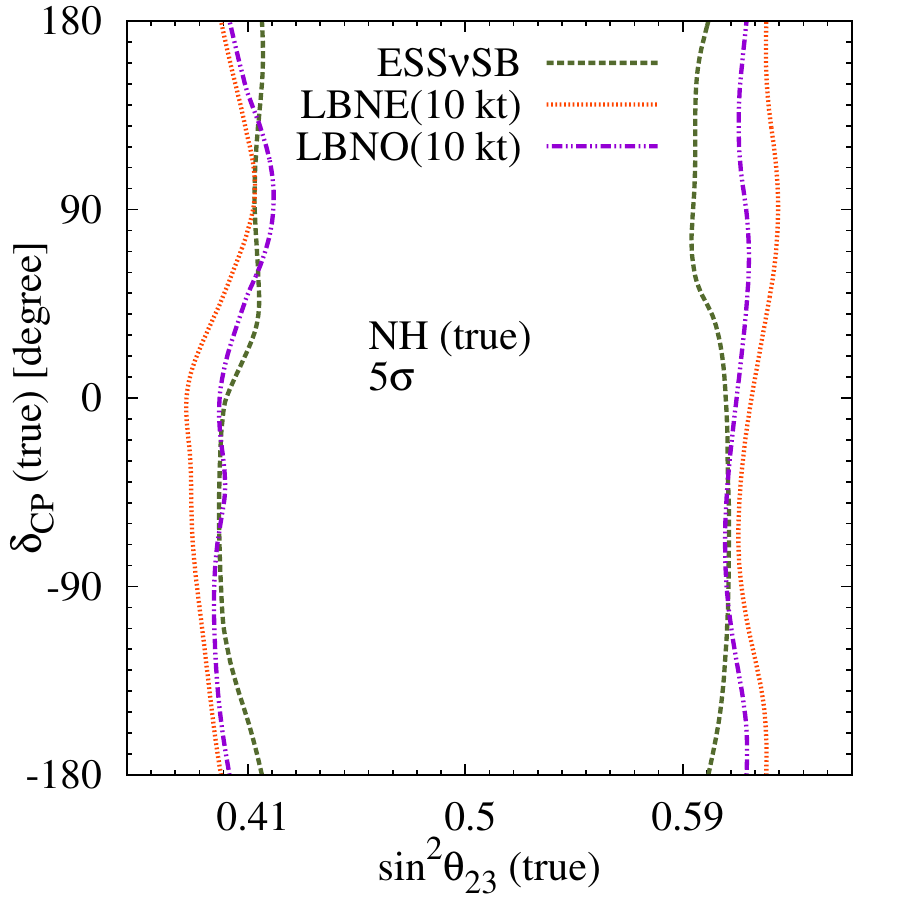}
\caption{\footnotesize{A comparison of the future facilities LBNE and LBNO with the 
ESS$\nu$SB set-up. {\bf Left panel: non-maximal $\tz$ discovery potential. Right panel:
octant resolution potential.} For ESS$\nu$SB, a 200 km long baseline and a $7\nu+3\anu$
running is considered. For both LBNE and LBNO, a 10 kt LArTPC detector and a
$5\nu+5\anu$ running is considered.  The horizontal black lines show 2\sig and 3\sig C.L.}}
\label{novat2kfigures}
\end{figure}

Finally, we end this paper with a comparison of the 
deviation from maximality and octant of 
$\theta_{23}$ discovery reach of the ESS$\nu$SB set-up with the 
other next-generation proposed long baseline superbeam experiments. 
We show in Fig.~\ref{novat2kfigures} this comparison for the 
ESS$\nu$SB set-up with the 200 km baseline option and $7\nu+3\anu$ 
run-plan (green short dashed lines), LBNE with 10 kt 
liquid argon detector (orange dotted lines), and LBNO with 
10 kt liquid argon detector (purple dot-dashed lines). 
For LBNE and LBNO, we have used the experimental 
specifications as given in \cite{Agarwalla:2013hma}. 
In generating the plots for these three future facilities, 
we have added the projected data from 
T2K ($2.5\nu+2.5\anu$) and NO$\nu$A ($3\nu+3\anu$).
The details of these experiments are the same as considered
in \cite{Agarwalla:2012bv}.
The left hand panel of this figure shows the 
$\Delta \chi^2$ as a function of $\sin^2\tz$(true) for 
deviation of $\theta_{23}$ from its maximal value for 
$\dcp$(true)$=0$. The ESS$\nu$SB set-up is seen to 
perform better than the other two superbeam options, mainly 
due to larger statistics. With larger detectors, both LBNE and 
LBNO will start to be competitive. 
The right hand panel shows $5\sigma$ contours for the 
octant of $\theta_{23}$ discovery reach 
in the $\dcp$(true)-$\sin^2\tz$(true)
plane. The three experiments are very comparable, with the best 
reach coming for the ESS$\nu$SB set-up with the 200 km 
baseline option and $7\nu+3\anu$ run-plan. 

To conclude, among all four choices of the baselines, the best results for sensitivity
to deviation from maximality and resolution of octant is expected 
for the 200 km baseline option. 
On the other hand, 
chances for discovery of CP violation are best for 
the 540 km baseline, which 
sits on the second oscillation maximum and hence  
gives the maximum coverage in true $\dcp$. However, the 
CP violation discovery prospects 
for the 200 km baseline is only slightly worse. We have also seen 
that for all oscillation physics results, the 
$7\nu+3\anu$ run-plan provides the best sensitivity amongst the 
three run-plan choices considered. 
While we appreciate the merit of putting the detector 
at the second oscillation peak, this paper shows the advantage of 
another baseline option, in particular, 200 km.

\acknowledgments
{
We thank E. Fernandez-Martinez, L. Agostino, and T. Ekel\"{o}f for useful discussions. 
S.K.A acknowledges the support from DST/INSPIRE Research Grant
[IFA-PH-12], Department of Science and Technology, India. S.C. and S.P. acknowledge
support from the Neutrino Project under the XII plan of Harish-Chandra
Research Institute. S.C. acknowledges partial support from the European 
Union FP7 ITN INVISIBLES (Marie Curie Actions, PITN-GA-2011-289442).
}

\bibliographystyle{apsrev}
\bibliography{references}

\end{document}